\documentclass[acmtog, manuscript, screen]{acmart}

\usepackage{url}            
\usepackage{booktabs}       
\usepackage{multirow}       
\usepackage{subcaption}     
\usepackage{xcolor}         
\usepackage{listings}       
\usepackage{balance}        
\usepackage{enumitem}       
\usepackage{xspace}         
\usepackage{pifont}         
\usepackage{tcolorbox}      

\newcommand{\tool}[1]{\texttt{#1}}
\newcommand{\qwen}{\tool{Qwen Code}\xspace}

\newcommand{\swebench}{\textsc{SWE-bench}\xspace}

\newcommand{\eg}{\emph{e.g.,}\xspace}
\newcommand{\etal}{\emph{et~al.}\xspace}

\newtcolorbox{findingbox}[1][]{
  colback=gray!5,
  colframe=gray!50,
  fonttitle=\bfseries,
  title={#1},
  boxrule=0.5pt,
  arc=2pt,
  left=6pt, right=6pt, top=4pt, bottom=4pt,
}

\setcopyright{acmlicensed}
\acmJournal{TOSEM}
\acmYear{2026}
\acmVolume{0}
\acmNumber{0}
\acmArticle{0}
\acmDOI{}

\begin{document}

\title{Don't Blame the Large Language Model: How Agent Harness Evolution Shapes Coding Agent Quality}


\author{Oussama Ben Sghaier}
\orcid{0000-0003-2737-0952}
\affiliation{%
  \institution{Queen's University}
  \country{Canada}}
\email{oussama.sghaier@queensu.ca}

\author{Hao Li}
\orcid{0000-0003-4468-5972}
\affiliation{%
  \institution{Queen's University}
  \country{Canada}}
\email{hao.li@queensu.ca}

\author{Bram Adams}
\orcid{0000-0001-7213-4006}
\affiliation{%
  \institution{Queen's University}
  \country{Canada}}
\email{bram.adams@queensu.ca}

\author{Ahmed E. Hassan}
\orcid{0000-0001-7749-5513}
\affiliation{%
  \institution{Queen's University}
  \country{Canada}}
\email{ahmed@cs.queensu.ca}

\begin{abstract}
Coding agents, autonomous systems that use large language models (LLMs) to resolve software engineering tasks, rely on \emph{agent harness}: a middleware layer in between a developer and a large language model that orchestrates system prompts, tool execution, context management, and iterative reasoning loops. 
While these agent harnesses evolve at extreme velocities, no study has examined how this evolution affects agent quality (i.e., effectiveness and efficiency) over time. Practitioners regularly report quality regressions after agent harness updates, yet consistently attribute them to the underlying model rather than the harness itself. 

In this paper, we address this gap by conducting the first controlled longitudinal study that isolates the agent harness contribution. Unlike prior work that fixes the agent harness and varies the model, we fix the model and vary only the agent harness, evaluating 35 sequential releases to measure their impact on agent effectiveness and efficiency.
We first empirically study the development and release evolution of five major open-source agent harnesses (i.e., Codex, Qwen Code, Gemini, OpenCode, and OpenHands), revealing extreme release velocities exceeding two releases per day and thousands of issues within months. We then perform a controlled deep dive into 35 sequential releases of the Qwen Code CLI, evaluating each against 50 stratified \swebench~Verified tasks while holding the underlying LLM constant. We trace the resulting quality fluctuations to specific development patterns and architectural components, and illustrate our findings with concrete qualitative evidence linking individual pull requests to measured quality shifts.

Our findings reveal that despite continuous development activity and growing codebase complexity of the agent harnesses, there is no statistically significant improvement in \swebench~benchmark score (i.e., resolve rates of bugs) across releases for a given fixed LLM version. Worse, later agent harness versions consume nearly double the computational tokens and tool calls without corresponding quality gains. We explain this paradox from two angles: at the \emph{project level}, we identify the development patterns (e.g., feature additions, fix-heavy releases, scattered small changes) correlating with quality fluctuations, while at the \emph{architecture level}, we localize regressions to specific high-risk architectural components. Our findings call for a new practice of quality assurance in the development of agent harnesses.

\end{abstract}



\maketitle


\section{Introduction}
\label{sec:introduction}

Software engineering is undergoing a fundamental transformation. AI-assisted development has progressed from inline code completions and code generation~\cite{vaswani2017attention, brown2020language, chen2021codex} to a new paradigm, SE~4.0 (i.e., Agentware)~\cite{hassan2024rethinking}, in which autonomous, goal-driven agents collaborate with human developers as AI teammates~\cite{hassan2025sase}. These coding agents no longer merely suggest code. They navigate codebases, plan multi-step changes, invoke compilers and test suites, and submit pull requests with minimal human oversight. The scale is already substantial: Li~\etal~\cite{li2025aidev} documented over 456,000 pull requests authored by five leading AI coding agents across more than 61,000 repositories over a six-month period. These agents are active participants in real-world software development, operating at a scale and velocity that would have been inconceivable years ago.

Behind every coding agent lies a critical but often overlooked software layer: the \emph{agent harness}. Agents such as Claude Code, Gemini CLI, Codex, and Qwen Code all require such non-model middleware, which manages interactions with an LLM by taking care of system prompts, tool definitions \cite{schick2023toolformer}, execution sandboxing \cite{wang2024openhands}, context management, and iterative reasoning loops \cite{sumers2023cognitive}.

\begin{figure}[t]
\centering
\includegraphics[width=1\columnwidth]{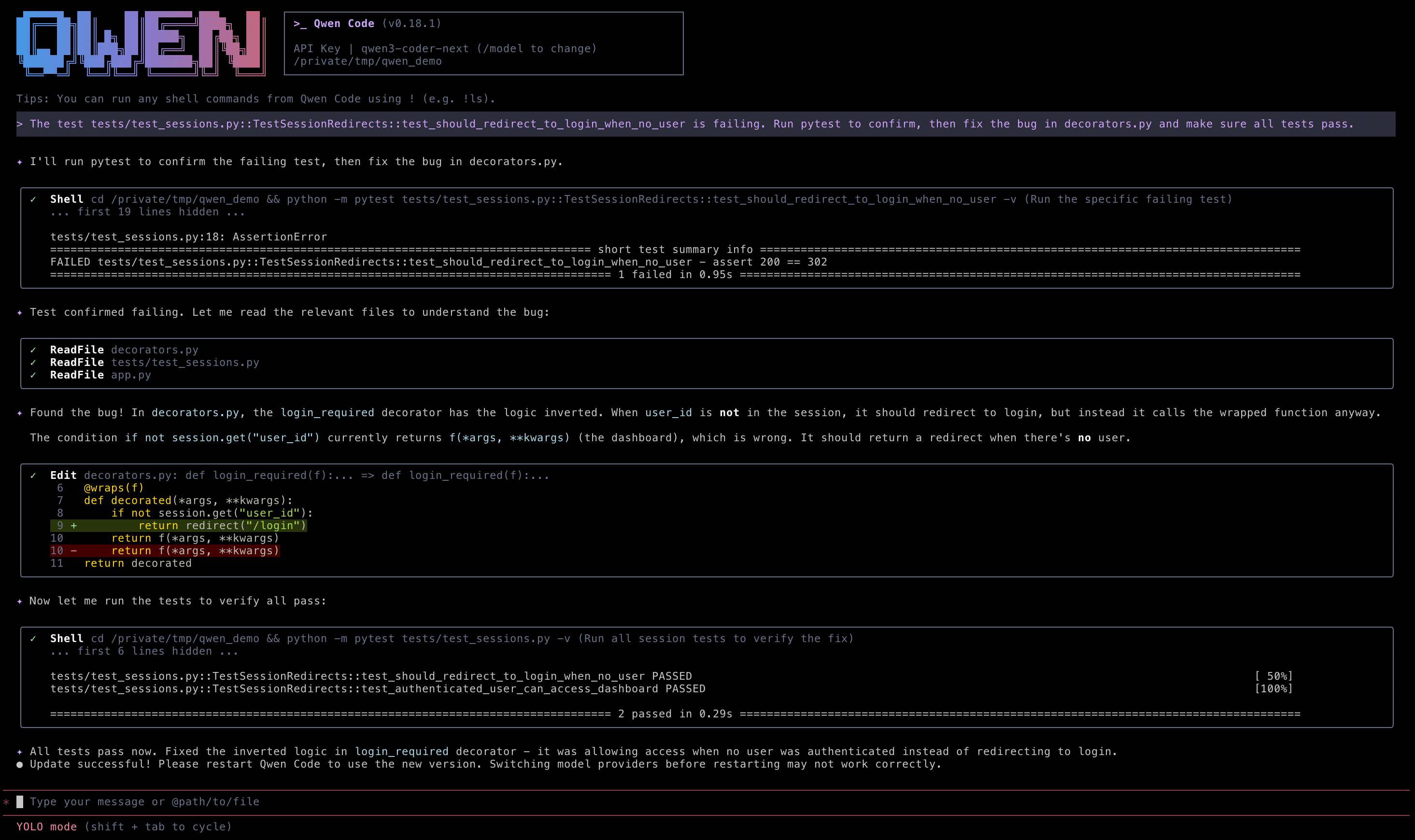}
\caption{An example interaction with Qwen Code v0.18.1 on a bug-fixing task. The agent iteratively reasons, invokes tools (\textsc{Shell}, \textsc{ReadFile}, \textsc{Edit}), observes results, and refines its plan until all tests pass.}
\label{fig:qwen-interaction}
\end{figure}

Figure~\ref{fig:qwen-interaction} illustrates a concrete interaction with \qwen~CLI, one of the agent harnesses studied in this paper. The agent receives a natural-language task (i.e., a failing test). In each turn, the LLM reasons about the next step (e.g., ``Test confirmed failing. Let me read the relevant files to understand the bug''), invokes a tool (e.g., \textsc{Shell} to run \texttt{pytest}, \textsc{ReadFile} to inspect source files, or \textsc{Edit} to apply a patch), and observes the result, which is appended to the conversation history and fed back to the model. The full conversation history, including the system prompt, all prior reasoning turns, tool calls, and their outputs, is prepended at every new LLM call. Finally, the figure also illustrates an agent harness update notification at the bottom that was applied without user intervention, underscoring how frequently these agents evolve and how silently those changes take effect.

The prevailing assumption among both practitioners and tool developers is that continuous agent harness improvement, e.g., adding layers of middleware for tool routing, state management, and multi-hop reasoning, leads to better agent quality. Each new release promises smarter context management, richer tool integration, or more sophisticated reasoning loops. Yet, practitioners regularly report that their agent's quality has deteriorated after updates, typically attributing this to the underlying model rather than the agent harness layer.\footnote{%
  Cursor version regression: \url{https://forum.cursor.com/t/cursor-is-getting-worse-and-worse/66070}\\
  Claude Code token consumption spike: \url{https://github.com/anthropics/claude-code/issues/16856}\\
  Claude Code Opus 4.6 quota burn: \url{https://github.com/anthropics/claude-code/issues/23706}\\
  Gemini CLI quality degradation: \url{https://github.com/google-gemini/gemini-cli/discussions/9073}\\
  Codex harness change causing quality cliff: \url{https://github.com/openai/codex/issues/8272}\\
  Qwen Code excessive token consumption: \url{https://github.com/QwenLM/qwen-code/issues/87}%
} In rare cases, some users recommend pinning an older agent harness version when they notice such regressions. Yet, the role of the agent harness layer itself remains largely uninvestigated. No study has empirically examined this disconnect.

LLMs themselves have received substantial research attention in software engineering. Prior work has evaluated their coding capabilities using benchmarks such as HumanEval~\cite{chen2021codex} and SWE-bench~\cite{jimenez2024swebench}, investigated reasoning and tool-use strategies through approaches such as Chain-of-Thought~\cite{wei2022chain} and ReAct~\cite{yao2023react}, and explored their application to a wide range of software engineering tasks~\cite{hou2024llm4se, liu2024llmagent}. More recently, agent harnesses such as SWE-agent~\cite{yang2024sweagent}, OpenHands~\cite{wang2025openhands}, and Agentless~\cite{xia2024agentless} have demonstrated that components beyond the underlying model, including tool interfaces, context management mechanisms, and execution workflows, can play a critical role in agent quality. However, existing studies~\cite{openhands2026index,  wong2025confucius} typically treat these agent harnesses as fixed infrastructure and focus on comparing different foundation models within the same agent harness. Consequently, while we increasingly understand how different models behave within a given agent harness, we know far less about how the agent harness itself evolves over time and how changes to that agent harness affect agent behavior, efficiency, and reliability~\cite{hasan2026empirical, rombaut2026inside}.

In this paper, we conduct a longitudinal empirical study in which we keep the underlying LLM version fixed and vary only the agent harness. We examine how the continuous evolution of agent harnesses affects agent effectiveness and efficiency (i.e., resolve rate, token consumption, and tool calls) and what development patterns and architectural factors are associated with these quality shifts.
We establish the scale and velocity of agent harness evolution across multiple coding agent harnesses, and perform a longitudinal controlled evaluation of 35 sequential releases of \qwen~CLI. \qwen~CLI is a fork of Gemini CLI (a widely adopted agentic CLI) and is built to support the open-source Qwen model family. We select \qwen~CLI because it is a high-performing, actively developed agent harness that natively supports custom local LLM endpoints, enabling us to hold the model constant across all releases. Our study is guided by four research questions (RQs):

\noindent\textbf{RQ0: What are the scale and evolutionary characteristics of coding agent harness projects?}

\noindent\textbf{Motivation.} Reasoning about how agent harness evolution affects quality first requires understanding the pace and scale of that evolution. Coding agent harnesses are developed by major AI companies as primary products built to showcase their own models, and they have attracted massive adoption within months of release. Prior work has characterized release patterns in traditional open-source software~\cite{khomh2015rapid, khomh2012faster}, but no study has systematically quantified the evolutionary characteristics of coding agent harness, including release velocity, development activity, and maintenance burden. Establishing this baseline provides an overview of the landscape of these emerging software systems, characterizes what makes their development practices distinctive, and motivates our subsequent research questions.

\noindent\textbf{Findings.}
Coding agent harness projects exhibit a development intensity that far exceeds that of traditional open-source projects. The five coding agents in our study average between 1.5 and 18 releases per week and 2.8--34 commits per day, with median PR review times under four hours, compared to 0.6--0.8 releases per week and 4.8--47.9 commits per day for VSCode and GitHubCLI over the same period. These agents accumulate thousands of issues within months of launch and struggle to close them at the same rate, leading to growing backlogs. Bug-fix commits account for roughly 30\% of all development effort, reflecting a reactive development cycle in which stabilization and feature development proceed simultaneously. Taken together, the agent harness surrounding the LLM is not stable but is instead a continuously evolving software system. This raises the question of whether this rapid evolution translates into actual quality improvements, in terms of effectiveness (i.e., correctness of the task) and efficiency (e.g., token consumption, tool call overhead). We term this pattern \emph{hyper-churn}; an unprecedented release velocity and development intensity that far exceeds that of traditional open-source software.

\noindent\textbf{RQ1: How does agent quality evolve across successive agent harness releases?}

\noindent\textbf{Motivation.} The central assumption driving agent harness development is that continuous improvement (e.g., richer tool integration, smarter context management, more sophisticated reasoning loops) translates into better agent quality. Yet, users frequently report on public forums and issue trackers that upgrading to a newer version degrades their agent's behavior, and attribute these regressions to the underlying model rather than the agent harness. Empirically verifying whether agent harness evolution actually improves quality or at least impacts it requires holding the model constant across agent harness releases. Otherwise, any observed change in quality conflates model updates with quality changes. No prior study has conducted such a controlled longitudinal evaluation. We address this gap by evaluating 35 sequential releases of \qwen~CLI on 50 stratified tasks from \swebench~Verified~\cite{jimenez2024swebench} while keeping the underlying LLM fixed, directly isolating the agent harness impact on quality.

\noindent\textbf{Findings.} Despite substantial growth in codebase size and continuous development activity, there is no statistically significant improvement in task resolve rates across the 35 \qwen~releases. Early versions of the agent harness sometimes even outperform their more sophisticated successors. Moreover, later releases consume significantly more computational resources, with token usage and tool call counts more than doubling in some cases, without corresponding gains in task resolution. This reveals a fundamental disconnect between development activity and agent effectiveness.

\noindent\textbf{RQ2: What release-level development patterns are associated with quality shifts?}

\noindent\textbf{Motivation.} Knowing that quality shifts occur across agent harness releases is not sufficient for practitioners, they need to understand which development patterns drive them. Prior work on traditional software has linked release characteristics such as code churn, commit composition, and contributor activity to software quality outcomes~\cite{khomh2015rapid}. However, it is unclear whether these relationships hold for agent harness, where behavioral quality is emergent and not captured by conventional quality metrics. We analyze the code changes, commit type distributions, contributor activity, and release size characteristics across all 35 releases to identify patterns that distinguish quality-improving releases from quality-degrading ones.

\noindent\textbf{Findings.} Feature-heavy agent harness releases are significantly correlated with higher resolve rates ($\rho=0.438$) at the cost of increased token consumption and tool calls. Fix-heavy releases are associated with higher token consumption without improving resolve rates. Releases with larger average PR sizes correlate with lower token consumption, suggesting that consolidating changes into coherent PRs reduces overhead relative to many small PRs. Removing code correlates with cost reductions, as higher code deletion volume is associated with lower token consumption.

\noindent\textbf{RQ3: Which architectural components of agent harness exhibit greater sensitivity to change?}

\noindent\textbf{Motivation.} The release-level patterns of RQ2 explain when quality shifts occur, but not where within the agent harness they originate. Agent harness is a complex system with multiple interacting components, including LLM communication layers, context management, tool execution, and security sandboxing. Changes to different agentic components carry different regression risks, but this has never been systematically characterized. We map all commits across the 35 releases to a reference agent harness architecture and analyze whether specific modifications to each architectural component are associated with quality degradation or improvement.

\noindent\textbf{Findings.} Not all architectural components carry equal quality regression risk. The \emph{LLM Provider} layer and \emph{Context Management} emerge as high-risk zones. Modifications to these components are most frequently associated with quality degradation, likely because they directly govern how information is passed to and from the model. By contrast, changes to \emph{Extensibility} and \emph{Security} components are consistently associated with safe or neutral quality outcomes, making them lower-risk components.

Across RQ1–RQ3, we find that agent harness complexity, token consumption, and tool calls increase significantly while benchmark quality (i.e., the ability to fix bugs) does not. Feature-heavy releases drive short-term effectiveness gains at the cost of efficiency, and specific architectural components carry disproportionate regression risk. These findings share a common underlying cause, which is the absence of Agentic Quality Assurance, i.e., automated quality regression testing that evaluates the agent's actual non-functional quality (e.g., token consumption, tool calls overhead) rather than merely verifying code correctness of a generated patch. The quality regressions we document are emergent behavioral effects of the interaction between the agent harness and the LLM, and they are not catchable by conventional unit or integration tests. Our inspection of the project's CI/CD pipelines reveals that every concrete example of quality degradation we document passed all existing automated checks, confirming that current testing practices of modern agent harnesses lack the mechanisms necessary to prevent agentic quality regressions, likely because of the cost of exhaustively running benchmark evaluations in CI/CD pipelines. 

Our findings carry practical implications for both agent harness developers and researchers. Developers should treat agent harness as quality-critical software that requires evaluation beyond functional correctness. Researchers evaluating coding agents should report and control for the agent harness version, not merely the underlying LLM, and should consider efficiency metrics (e.g., token consumption and tool calls) alongside effectiveness (e.g., resolve rate).

This paper makes the following contributions:

\begin{enumerate}[leftmargin=*]
    \item We characterize a phenomenon of ``hyper-churn'' across multiple coding agents, quantifying release velocity, development activity, and issue reports that significantly exceed those of mature classical open-source software engineering projects.
    \item We present the first longitudinal, controlled evaluation of 35 sequential agent harness releases of \qwen~on \swebench~Verified, isolating the harness impact on quality by holding the LLM constant.
    \item We provide project-level explanations of quality shifts by identifying and characterizing release-level development patterns, including code changes, commit types, and development velocity, that distinguish quality-improving releases from quality-degrading ones.
    \item We provide architecture-level explanations by constructing a component-level risk analysis, mapping commits to a standardized reference architecture to identify the architectural zones where modifications are most frequently associated with regressions.
    \item We provide a comprehensive replication package~\cite{replicationpackage2026} including all 35 \qwen~CLI versions, inference and evaluation scripts, and raw results.
\end{enumerate}


\section{Background}
\label{sec:background}

This section introduces the key concepts that underpin our study: coding agents, the agent harness layer that governs their behavior, and the benchmark we use for evaluation.

\subsection{Coding Agents}
\label{sec:bg:agents}

The evolution of AI-assisted software development has progressed through several stages~\cite{li2025aidev, hassan2025sase}. Early tools provided token-level predictions and inline code completions. The introduction of large language models (LLMs)~\cite{vaswani2017attention, brown2020language} elevated these tools to macro-level code generation, where models could produce entire functions or files from natural-language descriptions~\cite{chen2021codex}. The current frontier, often referred to as SE~3.0~\cite{hassan2025sase}, features \emph{coding agents}: autonomous systems that go beyond code generation to perform goal-directed software engineering tasks. These agents can read codebases, plan multi-step changes, invoke external tools (compilers, test runners, file editors), execute terminal commands, and iteratively refine their output based on feedback~\cite{yang2024sweagent, wang2024openhands}.

Coding agents are now active participants in real-world software development. Li~\etal~\cite{li2025aidev} documented over 456,000 pull requests authored by five leading agents (i.e., OpenAI Codex, Devin, GitHub Copilot, Cursor, and Claude Code) across more than 61,000 repositories over a six-month period. These agents operate as semi-autonomous collaborators, initiating pull requests, responding to review feedback, and engaging in iterative development loops with minimal human oversight.

\subsection{Agent Harness}
\label{sec:bg:scaffolding}

\begin{figure}[t]
\centering
\includegraphics[width=1\columnwidth]{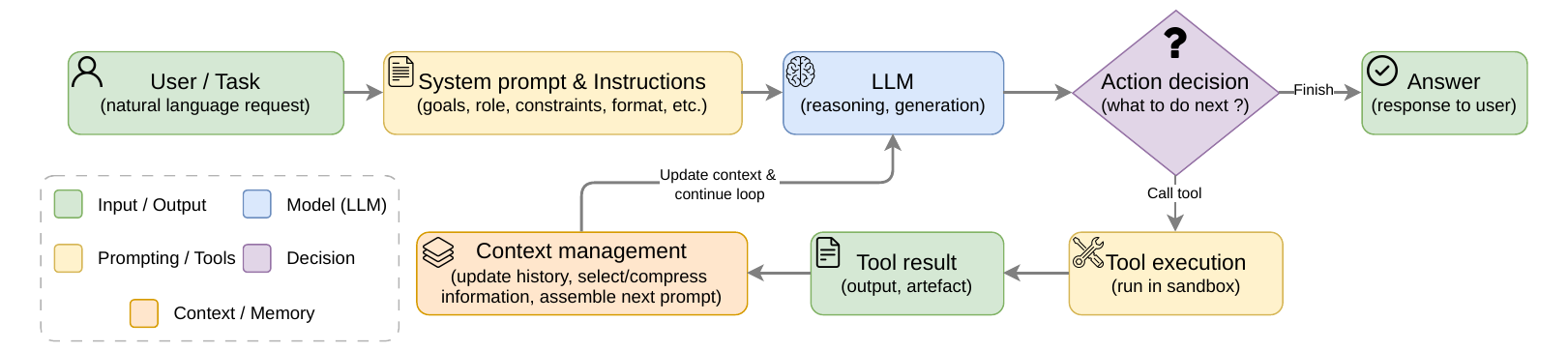}
\caption{Typical control flow within an agent harness. The agent harness orchestrates the iterative loop between the LLM and the execution environment.}
\label{fig:scaffolding-flow}
\end{figure}

While an LLM provides the core reasoning capability, a coding agent's behavior is largely governed by its \emph{harness}, also referred to as \emph{scaffold} (Figure~\ref{fig:scaffolding-flow}). It constitutes the software layer that mediates between the LLM and the execution environment~\cite{sumers2023cognitive, rombaut2026inside}. This layer shapes every aspect of the agent's operation: it defines the system prompts and instructions that specify the agent's goals, role, constraints, and response format~\cite{sumers2023cognitive}, specifies which tools the agent can invoke and how their results are fed back to the model~\cite{schick2023toolformer}, and manages the context provided to the LLM by updating history, selecting and compressing relevant information, and assembling the next prompt~\cite{sumers2023cognitive}.


As illustrated in Figure~\ref{fig:scaffolding-flow}, beyond shaping the LLM's inputs, the agent harness orchestrates the agent's execution loop. Through the action decision component, it determines whether the agent should produce a final answer or invoke external tools. Through tool execution, it enables the agent to safely execute commands in a sandboxed environment and obtain results~\cite{wang2024openhands}. Through context management, it incorporates tool outputs and assembles the next prompt, enabling the iterative reasoning cycle~\cite{yao2023react}. Each of the leading coding agents (e.g., OpenAI Codex and Claude Code) implements its own harness layer with different design choices for system prompt, tools, and context management~\cite{li2025aidev}.


An important but often overlooked aspect of agent harness is its impact on \emph{non-functional} quality attributes. Beyond functional correctness (does the tool execute?), the agent harness directly affects token consumption (how many LLM tokens are spent per task), tool call overhead (how many tool invocations the agent requires), and reasoning efficiency (whether the agent converges on a solution or gets trapped in unproductive loops). These non-functional properties have direct economic consequences, since cloud API pricing is tied to token usage, and practical consequences for user experience, since longer reasoning chains increase latency.

In practice, the agent harness and the LLM evolve largely independently, though some models (e.g., Claude, \qwen) are reportedly optimized during training for use with specific agent harnesses.\footnote{\url{https://github.com/QwenLM/qwen-code/tree/870bdf2a9d6fbd324d6d5d25735b2932dc4e402d\#why-qwen-code}}
Nevertheless, agent harness codebases evolve on a much faster timescale than model retraining cycles. An agent harness update can fundamentally alter agent behavior, changing which tools are available, how context is managed, or how the reasoning loop terminates, even when the underlying model remains unchanged. This separation is central to our study, where we hold the LLM constant across all 35 releases to isolate the agent harness contribution to observed quality variations.

\subsection{SWE-bench}
\label{sec:bg:swebench}

\swebench~\cite{jimenez2024swebench} is a widely adopted benchmark for evaluating the ability of coding agents to resolve real-world GitHub issues~\cite{yang2024swe, wang2024openhands, zhang2024autocoderover, xia2024agentless, comanici2025gemini, team2025kimi}. Each task consists of a real-world bug report, the corresponding repository, and a ground-truth patch validated by running the repository test suite. \swebench~Verified~\cite{chowdhury2024swebenchverified} is a curated subset of 500 tasks, filtered by human annotators to ensure each task is well-specified and reliably assessed. It spans popular Python repositories including Django, scikit-learn, Sphinx, SymPy, and Astropy. Tasks are stratified by difficulty based on the estimated time a human engineer would require to solve them, ranging from Easy ($<$15 min) to Medium (15 min--1 hr), Hard (1--4 hrs), and Very Hard ($>$4 hrs)~\cite{chowdhury2024swebenchverified}. The primary evaluation metric is the \emph{resolve rate}, i.e., the percentage of tasks for which the agent produces a patch that passes all associated tests.

\section{Methodology}
\label{sec:methodology}

\begin{figure*}[t]
\centering
\includegraphics[width=\textwidth]{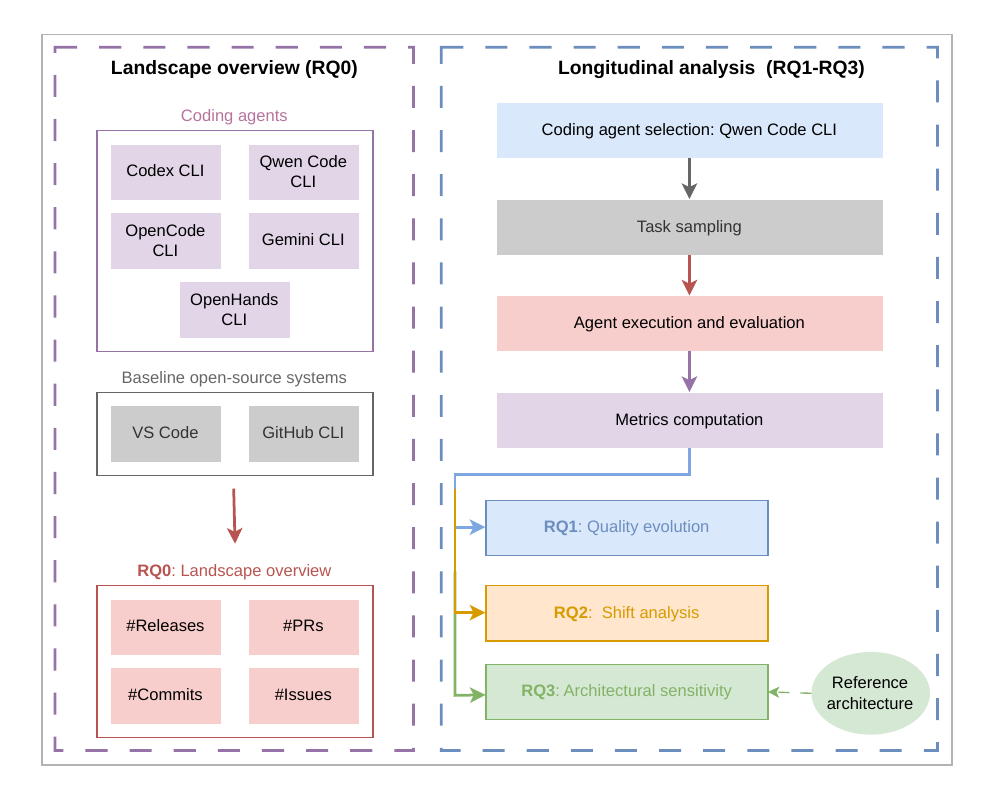}
\caption{Overview of our methodology. The landscape analysis (left) mines five coding agent harnesses and two baseline projects to establish a landscape overview of development velocity and maintenance burden (RQ0). The longitudinal analysis (right) selects Qwen Code for controlled longitudinal evaluation: 35 sequential releases are evaluated against 50 SWE-bench tasks using a fixed LLM (Qwen3-Next-80B-A3B), yielding quality evolution (RQ1), quality shift analysis (RQ2), and architectural sensitivity analysis (RQ3).}
\label{fig:methodology}
\end{figure*}

Figure~\ref{fig:methodology} illustrates the methodology that we follow in our study to address the RQs mentioned in the introduction. We begin with a broad landscape analysis across five coding agent harnesses, compared to two baseline projects, to characterize the high development activity and rapid release pace of these projects through releases, PRs, commits, and issues (RQ0). We then narrow our focus to a single agent (i.e., \qwen~CLI) for a controlled longitudinal evaluation, which follows four sequential steps: coding agent selection, task sampling, agent execution and evaluation, and metrics computation. These steps feed into three research questions.
RQ1 studies \emph{functional} and \emph{non-functional quality evolution} across 35 sequential releases using resolve rate, token consumption, and tool calls. RQ2 analyzes the potential relation of agent effectiveness and efficiency changes to 22 project-level development factors, including code churn, commit characteristics, and contributor activity. RQ3 conducts an \emph{architectural sensitivity analysis} to identify which components of the agent harness are most strongly associated with quality regressions and improvements.

As illustrated in Figure~\ref{fig:methodology}, the longitudinal analysis evaluates a single model (Qwen3-Next-80B-A3B-Instruct) across 35 sequential Qwen Code releases on 50 \swebench~Verified tasks, with two independent runs per task. While studying multiple agents in RQ0 ensures that the patterns we observe are not specific to a single project, focusing on a single agent in RQ1--RQ3 allows us to run controlled experiments (i.e., holding the LLM constant, evaluating every sequential release, tracing individual commits) that would not be feasible across all five agent harnesses.

\subsection{Coding Agent Selection}
\label{sec:method:subjects}

\subsubsection{Landscape overview: Coding Agent Harnesses (RQ0)}
\label{sec:method:subjects:broad}

For the landscape analysis, we focus on coding agent harnesses that are (1)~open-source with a publicly accessible Git history, (2)~actively maintained during our study period, and (3)~CLI-based, enabling automated evaluation without GUI dependencies. Among the systems satisfying these criteria, we select five widely used and well-documented agent harnesses: \textbf{Gemini CLI}~(Google), \textbf{Codex}~(OpenAI), \textbf{OpenCode}~(Anomaly), \textbf{OpenHands CLI}~(OpenHands), and \textbf{Qwen Code}~(Alibaba). These agent harnesses are chosen for their prominence, while providing sufficient release history to support longitudinal analysis, and have been used in recent coding agents research~\cite{zhang2026engineering, galster2026configuring, wang2025openhands, vangala2025ai, zhao2026specbench, nashid2025beyond}. Systems that are closed-source (e.g., Claude Code\footnote{Claude Code does maintain a public GitHub repository, but it contains only issue reports rather than the full source. Link to Claude Code GitHub repository: \url{https://github.com/anthropics/claude-code}}, GitHub Copilot), GUI-only (e.g., Devin), or with limited release activity are excluded, as they do not enable reproducible or temporal analysis of agent harness evolution. Table~\ref{tab:rq0-subjects} in Section~\ref{sec:rq0} summarizes each agent harness repository, language, lifespan, and scale.

To contextualize the development velocity, activity, and scale of coding agents against established software projects, we additionally select two mature, non-agentic open-source projects as baselines: \textbf{VS~Code} (Microsoft) and the \textbf{GitHub~CLI} (\texttt{cli/cli}). These projects represent different categories of widely adopted developer tools (a large-scale IDE and a command-line tool, respectively) with multi-year histories and conventional release cadences, making them suitable reference points for comparing release velocity, issue volume, and code churn.

\subsubsection{Longitudinal Analysis: Qwen Code CLI (RQ1--RQ3)}
\label{sec:method:subjects:deep}

For the longitudinal analysis, we select the \qwen~CLI, an agent harness that natively supports connection to a local OpenAI-compatible LLM endpoint. From its release history on npm, we identify 35 minor and patch releases (v0.0.10 through v0.10.3), excluding pre-release builds and nightly snapshots, enabling full control over the model version.
Earlier releases (eight stable versions: v0.0.1–v0.0.9) were excluded because they were released before \qwen~introduced native support for local OpenAI-compatible endpoints. This feature was added in v0.0.10.
Each release corresponds to a tagged commit in the Git repository, providing a complete and chronologically ordered release timeline. 

The CLI is designed and optimized specifically for the Qwen model family, meaning the agent harness is tailored to the model we use for evaluation. This makes it a natural pairing. The agent harness is expected to work well with the model, so any quality regressions we observe are attributable to the agent harness evolution rather than a mismatch between the agent harness and the model. 

We select Qwen3-Next-80B-A3B-Instruct, a state-of-the-art open-weight model widely adopted in the literature~\cite{harada2025automated, chen2026certain, hasan2026model}, as the largest Qwen model deployable on our dedicated server. This model is part of the Qwen3 family, which is supported across major inference engines and coding-agent harnesses and is regularly evaluated on major public leaderboards and benchmarks for code generation and agentic tasks~\cite{yang2025qwen3, cao2026qwen3}.
Hosting this model locally via vLLM allows us to hold the model version constant across all 35 releases, ensuring a fully controlled experimental design.
The CLI also implements all ten components of our reference architecture (Section~\ref{sec:method:architecture}), making it suitable for architectural sensitivity analysis. Finally, 35 releases is a manageable number that allows us to evaluate every single release (35 releases $\times$ 50 tasks $\times$ 2 runs = 3500 inference runs, plus evaluation).

\subsection{Task Sampling}
\label{sec:method:tasks}

We draw our evaluation tasks from \swebench~Verified, a curated subset of 500 real-world GitHub issues with test-validated ground-truth patches~\cite{jimenez2024swebench}. Prior to sampling, we excluded 8 tasks from the dataset whose evaluation environments failed even under the gold patch.\footnote{\url{https://github.com/SWE-bench/SWE-bench/issues/354}} From the remaining pool, we utilize the \swebench~Verified difficulty levels (estimated based on the time a human engineer required to fix the issue~\cite{chowdhury2024swebenchverified}) for stratified sampling to select a representative 10\% subset (50 tasks). Table~\ref{tab:task-sampling} shows the final distribution. 

Tasks were sampled proportionally across four difficulty levels, i.e., easy (20 tasks, $<$15 min fix time), medium (25 tasks, 15 min–1 hr), hard (5 tasks, 1–4 hrs), and very hard (0 tasks, $>$4 hrs, representing only 0.6\% of the pool). This ensures representative coverage of the full difficulty spectrum.

\begin{table}[t]
\caption{Stratified task sampling from \swebench~Verified.}
\label{tab:task-sampling}
\centering
\begin{tabular}{lrrr}
\toprule
\textbf{Difficulty} & \textbf{Fix Time} & \textbf{SWE-bench \%} & \textbf{Sampled} \\
\midrule
Easy     & $<$ 15 min   & 38.8\% & 20 (40\%) \\
Medium   & 15 min--1 hr & 52.2\% & 25 (50\%) \\
Hard     & 1--4 hrs     & 8.4\%  & 5 (10\%)  \\
Very Hard & $>$ 4 hrs   & 0.6\%  & 0 (0\%)   \\
\midrule
\textbf{Total} & & & \textbf{50} \\
\bottomrule
\end{tabular}
\end{table}

\subsection{Agent Execution and Evaluation}
\label{sec:method:infrastructure}

This step has three stages: model serving, agent execution, and patch evaluation.

\paragraph{Model Serving.} We host \textbf{Qwen3-Next-80B-A3B-Instruct}, a Mixture-of-Experts model widely used in the literature~\cite{harada2025automated, chen2026certain, hasan2026model}, with 80B total parameters (3B activated per token), on a dedicated server using vLLM~\cite{kwon2023vllm} (v0.18.0), exposed as an OpenAI-compatible API endpoint. We use the default sampling parameters of the \qwen~CLI. By self-hosting the model, we ensure that the LLM remains identical across all 35 releases, eliminating silent model updates as a confound.

\paragraph{Agent Execution.} Each release of the \qwen~CLI is installed from its corresponding npm package version and executed under Node.js~22. For each of the 50 sampled tasks, the agent receives the GitHub issue description and is given access to a checkout of the target repository (i.e., the open-source project containing the bug, as defined by the \swebench~Verified benchmark) at the relevant commit. We impose a timeout of 600 seconds per task to bound computational cost. Each task is run twice (run-1, run-2) to measure consistency.

\paragraph{Patch Evaluation.} We evaluate the agent-generated patches using the standard \swebench~Docker harness. Each patch, generated by the agent harness, is applied to the target repository, and the full test suite is executed in an isolated Docker container. A task is marked as resolved if and only if all tests pass. This ensures that evaluation is reproducible and independent of the host environment.

\subsection{Metrics Computation}
\label{sec:method:metrics}

We measure agent quality along two dimensions: \emph{effectiveness} (does the agent solve the task?) and \emph{efficiency} (how much resources does it consume?). To account for the non-deterministic nature of LLM generation, we execute each task twice (run-1 and run-2) and report all metrics as the average across both runs.

\begin{itemize}[leftmargin=*]
    \item \textbf{Resolve rate:} The percentage of tasks for which the agent produces a patch that passes all repository tests. This is the primary effectiveness metric, consistent with the standard \swebench~evaluation protocol. We calculate this by parsing the final test execution logs generated by the \swebench~Docker harness.
    
    \item \textbf{Token consumption:} Total number of input and output tokens consumed per task. This measures the computational cost of the agent's reasoning process. We compute this by aggregating the token usage metadata reported for every individual LLM call within the agent's structured trajectory execution logs generated by the \qwen~CLI. These logs record the agent's interactions, including LLM requests and responses, tool invocations, token usage statistics, and execution turns.
    
    \item \textbf{Tool call frequency:} The number of tool invocations per task. Combined with token consumption, this characterizes the agent's interaction pattern and resource intensity. We compute this by parsing the structured execution trajectory logs saved by the \qwen~CLI at the end of each session.
\end{itemize}

For RQ1, we use these metrics to categorize each release into one of three quality tiers (\emph{Good}, \emph{Neutral}, or \emph{Bad}) based on their distribution across all 35 releases. The categorization thresholds and statistical tests are described in Section~\ref{sec:rq1}.

\subsection{Reference Architecture}
\label{sec:method:architecture}

To systematically map code changes to architectural components (RQ3), we derive a \emph{reference software architecture} for the agent harness from the five agent harnesses in our study. In this context, a reference architecture captures the core components of agentic systems (e.g., planning, tool use, context management) and the interactions between them. It provides a common conceptual agent harness for analyzing and comparing different implementations of agent harnesses. Following the process established by Hassan and Holt~\cite{hassan2000reference}, we first derive a conceptual architecture for each agent harness from its documentation, then map it to a concrete architecture by tracing actual code components, iteratively refining both until they align. After repeating this process for all five agent harnesses, we compare the resulting concrete (aligned) architectures across projects to synthesize a unified ten-component reference architecture. Appendix~\ref{sec:architecture} presents the full derivation process, the architecture itself, and its validation through mapping to each agent harness concrete implementation.

This reference architecture serves as the basis for the architectural sensitivity analysis, as shown in Figure~\ref{fig:methodology}, where code changes to the \qwen~CLI harness are mapped to ten architectural components and correlated with observed quality shifts.

\section{RQ0: What are the scale and evolutionary characteristics of agent harness projects?}
\label{sec:rq0}


\subsection{Motivation}
Before we study how changes in agent harness affect quality, we first need to understand the general nature of these new software systems. AI coding agents are an emerging technology, so it is unclear if they follow the same development and release patterns as traditional software projects. Therefore, we want to explore how they are actually built and maintained in practice. By examining their scale, daily development activity, and release frequency, we can build a clear picture of how these projects evolve. This foundational understanding is necessary to contextualize our subsequent investigation into whether and why their quality might change over time.

\subsection{Study Design}
\label{sec:rq0:design}

We mine the GitHub repositories of five coding agent harnesses and two non-agentic baseline projects (Section~\ref{sec:method:subjects:broad}). Table~\ref{tab:rq0-subjects} summarizes the seven repositories. The agent harnesses are among the most prominent open-source coding agents available, collectively attracting over 280,000 GitHub stars and over 1,600 contributors. All five emerged within a span of seven months (April--November~2025).

The two baseline projects were chosen to represent different categories of mature, widely adopted developer tools. VS~Code (182.9K stars) is a large-scale IDE with a decade-long history and a well-established monthly release cadence, representing a complex, feature-rich desktop application. GitHub~CLI (43.3K stars) is a command-line tool (closer in form factor to the coding agents themselves) with a conventional bi-weekly release cycle. Together, these baselines span the spectrum from lightweight CLI tool to full IDE, providing robust reference points against which to measure the agent harnesses' development intensity. All metrics for baselines are computed over the same calendar period (March~2025--February~2026) as the agent harnesses, controlling for ecosystem-level effects and ensuring a fair comparison.

We analyze each project along four dimensions: release velocity, development activity (\#commits, \#PRs, and review times), maintenance burden (commit classification, \#issues, and issue backlog, defined as the net cumulative number of open issues awaiting resolution at the end of each month).
To compute these metrics, we mined the issue reports, pull requests, and commit histories of each repository using the GitHub API. Given the scale of the dataset (tens of thousands of commits), manual classification of commit types was infeasible. Instead, we leveraged the fact that these projects adhere to standard commit formatting conventions (e.g., the conventional commits specification\footnote{\url{https://www.conventionalcommits.org/}}). We automatically extracted the structured prefixes from each commit message (e.g., \texttt{fix:}, \texttt{feat:}, and \texttt{refactor:}) and mapped them to the corresponding categories defined in Hindle~\etal's taxonomy~\cite{hindle2008large}. Commits that lacked a standard prefix or did not fit a predefined category were classified as \emph{Other}.

\begin{table*}[t]
\caption{Overview of studied repositories. Activity metrics cover the study period (March~2025--February~2026). \#Stars as of February~2026.}
\label{tab:rq0-subjects}
\centering

\begin{subtable}{\textwidth}
\centering
\caption{Repositories' metadata}
\begin{tabular}{llll}
\toprule
\textbf{Harness} & \textbf{Repository} & \textbf{Lang.} & \textbf{First Release} \\
\midrule
Gemini CLI    & \texttt{google-gemini/gemini-cli}   & TS     & 2025-06-24 \\
Codex         & \texttt{openai/codex}               & Rust   & 2025-04-30 \\
OpenCode      & \texttt{anomalyco/opencode}         & TS     & 2025-05-14 \\
OpenHands CLI & \texttt{OpenHands/OpenHands-CLI}    & Python & 2025-11-18 \\
Qwen Code     & \texttt{QwenLM/qwen-code}           & TS     & 2025-08-01 \\
VS Code       & \texttt{microsoft/vscode}           & TS     & 2015-11-17 \\
GitHub CLI    & \texttt{cli/cli}                    & Go     & 2020-02-04 \\
\bottomrule
\end{tabular}
\end{subtable}

\vspace{0.6em}

\begin{subtable}{\textwidth}
\centering
\caption{Repositories' activity during the study period}
\begin{tabular}{lrrrrrrr}
\toprule
\textbf{Harness} & \textbf{\#Stars} & \textbf{\#Commits} & \textbf{\#Releases} & \textbf{\#Issues} & \textbf{\#PRs} & \textbf{\#Contributors} & \textbf{Lifespan (d)} \\
\midrule
Gemini CLI    & 93.6K  & 4,715  & 330 & 9,951  & 4,916  & 439 & 224 \\
Codex         & 61.0K  & 3,812  & 518 & 5,574  & 3,840  & 358 & 293 \\
OpenCode      & 106.7K & 9,521  & 716 & 8,621  & 2,420  & 454 & 278 \\
OpenHands CLI & 110    & 254    & 19  & 200    & 225    & 24  & 90 \\
Qwen Code     & 18.9K  & 3,990  & 288 & 1,029  & 472    & 327 & 201 \\
VS Code       & 182.9K & 16,855 & 40  & 40,074 & 11,438 & 324 & 352 \\
GitHub CLI    & 43.3K  & 1,721  & 33  & 1,118  & 389    & 384 & 362 \\
\bottomrule
\end{tabular}
\end{subtable}

\end{table*}

\subsection{Results}
\label{sec:rq0:results}


\paragraph{Finding 1: Agent harnesses release 13--28$\times$ more frequently than traditional open-source projects.}
The most striking characteristic of coding agent development is its release cadence. Table~\ref{tab:rq0-releases} presents release velocity metrics for the five agent harnesses. OpenCode leads with 18.0 releases per week (median interval of 0.12 days between releases), followed by Codex at 12.4/week and Gemini~CLI at 10.3/week. Qwen Code, the second youngest project, sustains 10.0 releases per week. Even OpenHands~CLI, which launched most recently (November~2025) and has the lowest velocity among the agents, still maintains a cadence of 1.5 releases per week, nearly twice that of the two baseline projects; i.e., 0.8 releases/week for VS Code, one of the most widely adopted IDEs in recent years~\cite{sergeyuk2025survey}, and 0.6 releases/week for GitHub CLI, a command-line tool of comparable form to the coding agent CLIs, making the most active agent harnesses 13--28$\times$ more release-intensive. 

\begin{table}[t]
\caption{Release velocity metrics across agent harnesses and baselines. ``Study Period'' counts releases during March~2025--February~2026; ``Lifetime'' counts all releases since the project launch. For agent harnesses, the two are identical since all launched within the study period.}
\label{tab:rq0-releases}
\centering
\begin{tabular}{lrrrrr}
\toprule
\textbf{Harness} & \textbf{\begin{tabular}[c]{@{}r@{}}\#Releases \\ (lifetime)\end{tabular}} & \textbf{\begin{tabular}[c]{@{}r@{}}\#Releases \\ (study period)\end{tabular}} & \textbf{\begin{tabular}[c]{@{}r@{}}Release frequency \\ ( per week)\end{tabular}}  & \textbf{\begin{tabular}[c]{@{}r@{}}Median Interval \\ between Releases\end{tabular}}  & \textbf{Patch/Minor Ratio} \\
\midrule
Gemini CLI    & 330 & 330 & 10.3 & 0.20 days & 4.9  \\
Codex         & 518 & 518 & 12.4 & 0.19 days & ---  \\
OpenCode      & 716 & 716 & 18.0 & 0.12 days & 39.1 \\
OpenHands CLI &  19 &  19 &  1.5 & 5.06 days & 1.3  \\
Qwen Code     & 288 & 288 & 10.0 & 0.59 days & 17.0 \\
\midrule
VS Code       &  200 & 40 &  0.8 & 6.99 days & 0.8  \\
GitHub CLI    &  191 & 33 &  0.6 & 11.31 days & 0.1  \\
\bottomrule
\end{tabular}
\end{table}

These results show that agent harnesses are undergoing rapid and continuous evolution. This rapid release cadence is not driven solely by occasional feature releases. Even within the same minor-version, harnesses undergo extensive iteration. For example, OpenCode averages 39.1 patch releases per minor version, while Qwen Code averages 17.0, compared to just 0.8 for VS Code and 0.1 for GitHub CLI. Although the precise meaning of version increments may vary across projects, particularly for pre-1.0 versions, these high patch-to-minor ratios indicate frequent updates and continuous refinement of the agent harness between major feature milestones. As a result, substantial cumulative changes can accumulate over relatively short periods of time.

\begin{figure}[t]
\centering
\includegraphics[width=0.8\columnwidth]{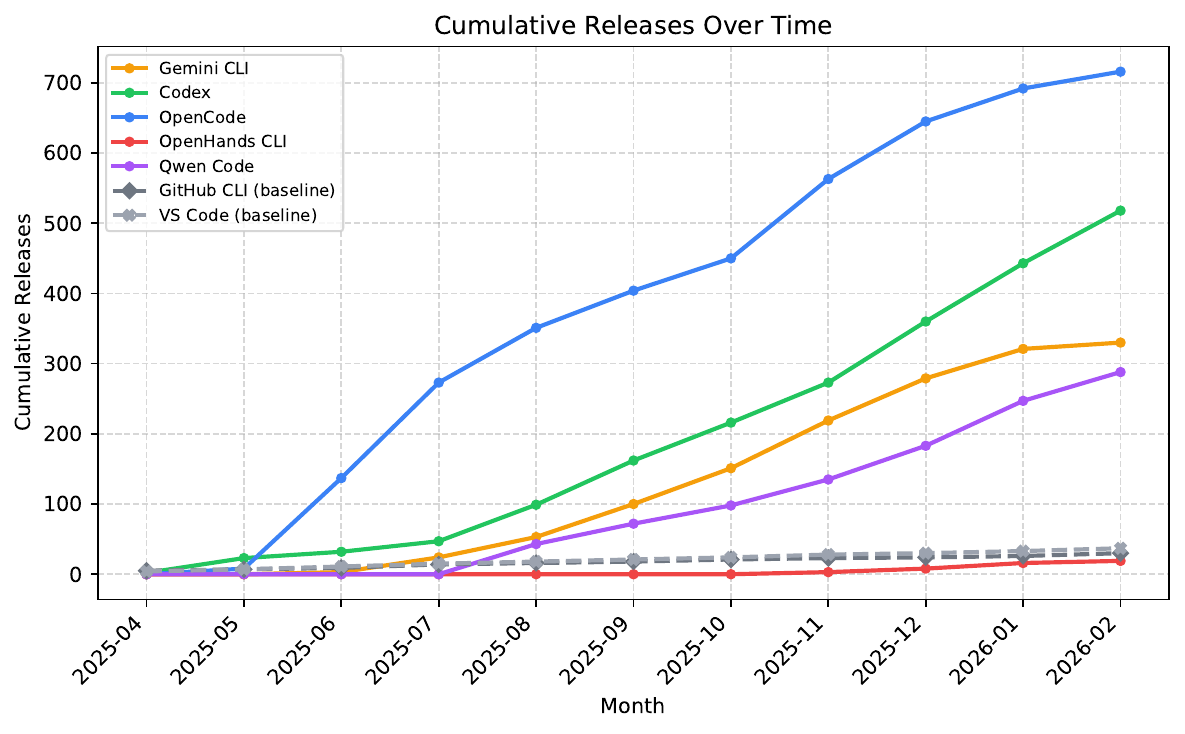}
\caption{Cumulative releases over time for the five agent harnesses and two baselines (dashed). The steep, sustained growth, particularly for OpenCode (716 releases in 278 days) and Codex (518 in 293 days), contrasts sharply with the near-flat baseline trajectories of VS~Code (40) and GitHub~CLI (33).}
\label{fig:rq0-cumulative-releases}
\end{figure}

The release cadence varies over time. Figure~\ref{fig:rq0-cumulative-releases} plots cumulative releases across the study period. Most agent harnesses settle into linear growth over time, with Gemini~CLI even showing a slowdown after its initial surge, though OpenCode and Codex maintain or accelerate their pace throughout the period. OpenCode peaked at 136 releases in a single month (July~2025), while Codex reached 87 (December~2025). 

\paragraph{Finding 2: Agent harnesses sustain 13--34 merged commits per day with near-instantaneous PR review times.}

Development activity mirrors the observed rapid release cadence. As shown in Table~\ref{tab:rq0-activity}, the four mature agent harnesses sustain between 13 and 34 merged commits per day, with OpenCode leading at 34.2 commits/day (9,521 total), followed by Gemini~CLI (21.0/day) and Qwen~Code (19.9/day). Pull request throughput is similarly high. Gemini~CLI merged 4,916 PRs in 224 days, while Codex merged 3,840 in 293 days. Despite comparable or even higher commit volumes in traditional projects (e.g., VS~Code at 47.9 commits/day), agent harnesses convert development activity into releases much more aggressively, requiring only 7--14 commits per release on average compared to 421.4 for VS~Code and 52.2 for GitHub~CLI. Furthermore, median PR review times remain below four hours for Gemini~CLI, Codex, and OpenCode, indicating that code changes move from development to release with minimal latency. 
These near-instantaneous PR review times may partly reflect the use of agentic AI for code review, given that these projects are themselves agent harnesses.
Together, these results reveal a development pipeline optimized for rapid iteration, where changes are merged, released, and deployed at a pace far exceeding that of traditional software projects.

\begin{table}[t]
\caption{Development activity metrics across agent harnesses and baselines.}
\label{tab:rq0-activity}
\centering
\begin{tabular}{lrrrrr}
\toprule
\textbf{Harness} & \textbf{\#Commits} & \textbf{\#Commits/day} & \textbf{\#Commits/release} & \textbf{\#PRs} & \textbf{Med. Review Duration} \\
\midrule
Gemini CLI    & 4,715  & 21.0 & 14.3 & 4,916 & 0.17 days \\
Codex         & 3,812  & 13.0 & 7.4 & 3,840 & 0.13 days \\
OpenCode      & 9,521  & 34.2 & 13.3 & 2,420 & 0.13 days \\
OpenHands CLI &   254  &  2.8 & 13.4 & 225 & 0.15 days \\
Qwen Code     & 3,990  & 19.9 & 13.8 & 472 & 0.77 days \\
\midrule
VS Code       & 16,855 & 47.9 & 421.4 & 11,438 & 0.03 days \\
GitHub CLI    &  1,721 &  4.8 & 52.2 & 389 & 1.63 days \\
\bottomrule
\end{tabular}
\end{table}



\paragraph{Finding 3: Agent harnesses accumulate issues faster than they can resolve them, leading to growing backlogs.}

Table~\ref{tab:rq0-issues} summarizes reported issue volume and resolution metrics across all projects. Gemini~CLI accumulated 9,951 issues in 224 days (44.4/day), OpenCode received 8,621 in 278 days (31.0/day), and Codex attracted 5,574 in 293 days (19.0/day), rates that far exceed those of even heavily used traditional projects. VS~Code received 40,074 issues over the same 12-month period (109.8/day), but with a close rate of 89.3\% and a dedicated team averaging 68 active contributors per month. GitHub~CLI, a command-line tool of comparable form factor to the coding agent CLIs, received just 1,118 issues (3.1/day) with a close rate of 82.1\%.


Issue resolution rates reveal further maintenance challenges. OpenCode closes only 54.0\% of its issues, leaving a backlog of 3,965 open issues. Gemini~CLI performs better at 81.8\%, similar to VS~Code (89.3\%) and GitHub~CLI (82.1\%).
Median resolution time also differs. OpenCode resolves issues in a median of 0.82 days, while Gemini~CLI takes 10.84 days and Qwen~Code takes 6.21 days. The fast resolution time for OpenCode, combined with its lower close rate, suggests a triage strategy that quickly addresses the most solvable issues while deferring the rest.

\begin{table}[t]
\centering
\caption{Issue volume and resolution metrics across coding agent harnesses and baseline projects during the studied period.
Close rate = \#closed / \#total issues.}
\label{tab:rq0-issues}
\begin{tabular}{l r r r r r}
\toprule
Project & \#Total & \#Open & \#Closed & Close Rate & \#Issues/Day \\
\midrule
Gemini CLI  & 9,951 & 1,808 & 8,143 & 81.8\% & 44.4 \\
OpenCode    & 8,621 & 3,965 & 4,656 & 54.0\% & 31.0 \\
Codex       & 5,574 & 1,238 & 4,336 & 77.8\% & 19.0 \\
Qwen Code   & 1,029 &   444 &   585 & 56.9\% & --- \\
OpenHands   &   200 &    62 &   138 & 69.0\% & --- \\
\midrule
VS Code     & 40,074 & 4,278 & 35,796 & 89.3\% & 109.8 \\
GitHub CLI  &  1,118 &   200 &    918 & 82.1\% & 3.1 \\
\bottomrule
\end{tabular}
\end{table}

\begin{figure}[!htbp]
\centering
\includegraphics[width=0.8\columnwidth]{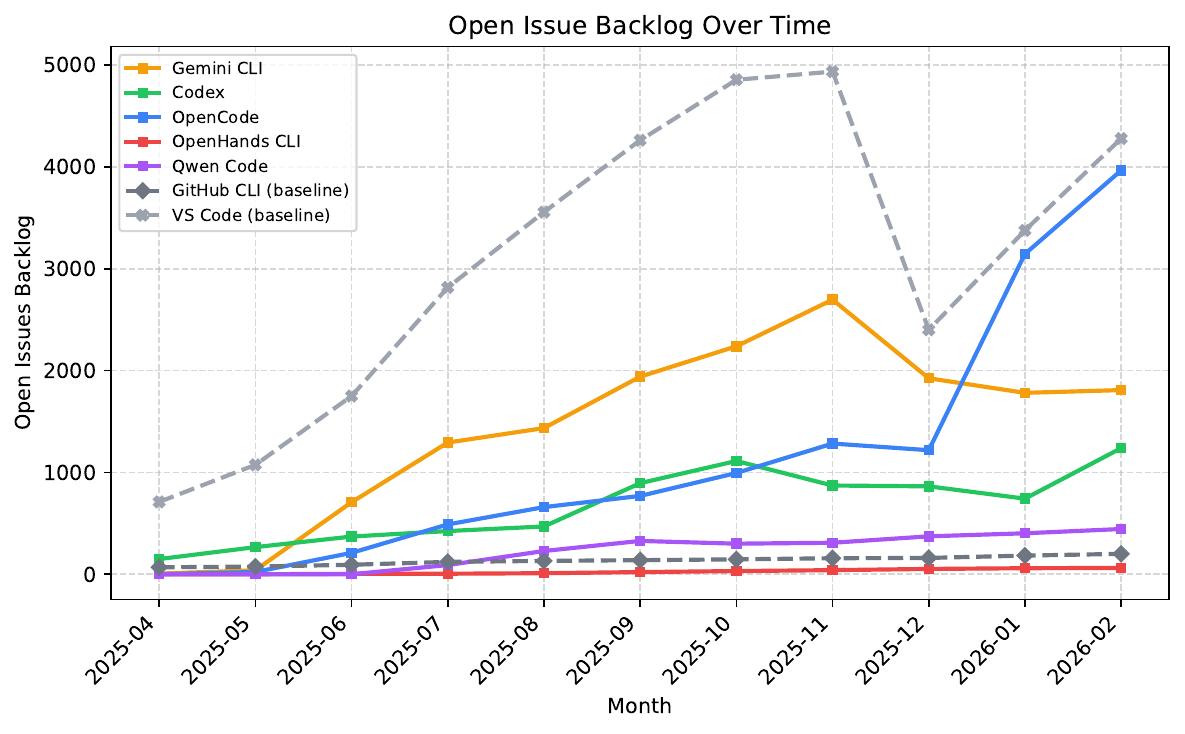}
\caption{Net cumulative number of open issues awaiting resolution~(i.e., backlog) at the end of each month over time. OpenCode's backlog grows steeply through January~2026, while Gemini~CLI stabilizes after an initial surge. The baselines (dashed) remain comparatively flat, with GitHub~CLI's backlog under 200 throughout the period.}
\label{fig:rq0-issue-backlog}
\end{figure}

Figure~\ref{fig:rq0-issue-backlog} plots the open issue backlog over time for all seven projects. It is defined as the net cumulative number of open issues awaiting resolution at the end of each month. OpenCode's backlog grew from zero to nearly 4,000 open issues in under a year, with a sharp spike in January~2026 when 3,103 new issues were opened in a single month. Gemini~CLI's backlog stabilized around 1,800 after an initial surge, suggesting a more effective triage process. The baselines tell a strikingly different story where GitHub~CLI's backlog grew only to 200 open issues over the entire period, and even VS~Code, despite receiving 40,074 issues, kept its backlog between 700 and 4,900 through aggressive closure. The sharp drop in VS Code's backlog observed in December 2025 corresponds to an end-of-year housekeeping,\footnote{\url{https://code.visualstudio.com/updates/v1_108}} during which maintainers and automated bots closed 5,951 issues in a single month to triage stale reports.
These trajectories reveal that issue volume is not merely a function of project popularity; it reflects the interaction between release frequency, user adoption, and maintenance capacity.

\begin{findingbox}[Summary of RQ0]
Coding agent harnesses exhibit a development intensity that far exceeds that of traditional open-source projects, averaging 10--18 releases per week, 13--34 merged commits per day, and median PR review times under 4 hours, compared to 0.6--0.8 releases per week for VSCode and GitHub CLI. These harnesses accumulate thousands of issues within months.
\end{findingbox}

\section{RQ1: How Does Agent Quality Evolve Across Successive Releases?}
\label{sec:rq1}

\noindent

\subsection{Motivation}

RQ0 established that agent harness projects evolve at extreme rates, with 10--18 releases per week and thousands of commits per month. Crucially, most coding agent CLIs are configured to auto-update or heavily prompt users to upgrade, silently pulling developers onto these new releases unless they explicitly opt out. This persistent exposure to rapid change raises a natural follow-up question: does this evolution translate into measurable benchmark quality improvements for the agent? If continuous agent harness development does not improve task resolution, token efficiency, and/or tool call overhead, then the high development intensity documented in RQ0 may be producing complexity without value. To answer this question in a controlled manner, we fix the LLM version and vary only the agent harness version across 35 sequential releases.

\subsection{Study Design}
\label{sec:rq1:setup}

\subsubsection{Agent Harness Selection and Version Control}

We evaluate 35 sequential minor and patch releases\footnote{Version v0.0.11 is excluded due to a known runtime bug that severely restricted tool availability, rendering its traces non-representative.} of the \qwen~CLI, spanning v0.0.10 through v0.10.3. Earlier versions (prior to v0.0.10) were excluded because they lacked logging instrumentation or could not connect to local OpenAI-compatible server endpoints. To isolate the agent harness as the sole independent variable, we hold the LLM version constant across all 35 releases and automated the environment setup by sequentially installing each target version via \texttt{npm install}.

\subsubsection{Local Model Configuration}

To ensure that observed quality fluctuations are attributable solely to agent harness changes rather than model updates or stochasticity, we held the LLM version constant across all 35 releases as described in Section~\ref{sec:method:infrastructure}. Qwen3-Next-80B-A3B-Instruct is a state-of-the-art open-weight model that is widely adopted by practitioners, making it a natural pairing with \qwen~CLI, which is designed and optimized specifically for the Qwen model family.

\subsubsection{Evaluation Methodology and Metrics}
\label{subsubsec:methodology-metrics}
We evaluated each release against a stratified subset of 50 tasks from \swebench~Verified~\cite{jimenez2024swebench}, ensuring exposure to diverse real-world bug complexities while keeping computational costs tractable. To account for stochastic execution noise, we ran each task twice per agent harness version (35 versions $\times$ 50 tasks $\times$ 2 runs = 3,500 total executions).


For each task execution, we extract two categories of quality metrics. To measure \textbf{efficiency}, we parse the \qwen~CLI's structured execution logs, which record every tool invocation and token consumption throughout the agent's reasoning trajectory. To measure \textbf{effectiveness}, we run the agent's final generated patch through \swebench~evaluation, which executes the target repository's full test suite within an isolated Docker container and determines whether the patch resolves the issue. From these two sources, we derive three metrics: \textbf{token consumption} (total input/output tokens per task), \textbf{\#tool calls} (total tool invocations per task), and \textbf{resolve rate} (percentage of tasks where the generated patch passes all tests).
We also tracked the number of \textbf{conversation turns}, defined as the back-and-forth reasoning roundtrips between the agent harness and the LLM during execution.

To capture execution variance without masking it behind a single aggregate, we report metrics independently for Run1 and Run2 alongside their combined average. Resolve Rate is the unweighted mean of the two run-level success rates. Token Consumption and Tool Calls are the average of the two run-level per-task means. Reporting both runs individually allows us to visually confirm that observed trends are systematic across trials rather than artifacts of a single noisy execution.
We use the Wilcoxon signed-rank test~\cite{conover1999practical} to compare metric distributions between Run~1 and Run~2, and Spearman rank correlation~\cite{spearman1904proof}, which operates on ranks rather than raw values, to assess monotonic trends of quality metrics (i.e., resolve rate, token consumption, tool calls) across releases.

Some tasks are inherently harder and consistently consume more tokens and require more tool calls regardless of the agent harness version. A version that successfully attempts more difficult tasks may appear more expensive simply due to task selection, not agent harness behavior. To remove this task-difficulty bias, we normalize both metrics at the task level. For each task~$t$, we compute a baseline $\bar{m}_t$ as that task's mean metric value across all 35 versions and both runs. Each version's value for that task is then expressed as a percentage deviation from this baseline: $\frac{m_{t,v} - \bar{m}_t}{\bar{m}_t} \times 100$. A value of $+20\%$ means the agent harness version consumed 20\% more tokens (or issued 20\% more tool calls) than that task's typical cost. A value of $-20\%$ means it consumed 20\% less. Averaging these deviations across all tasks yields a per-version efficiency signal that reflects the agent harness's own contribution to resource consumption (i.e., token consumption and tool calls), independent of which tasks happened to be harder or easier to solve.

\subsection{Results}
\label{sec:rq1:results}

\begin{figure*}[!htbp]
\centering
\begin{subfigure}[b]{\textwidth}
    \centering
    \includegraphics[width=\textwidth]{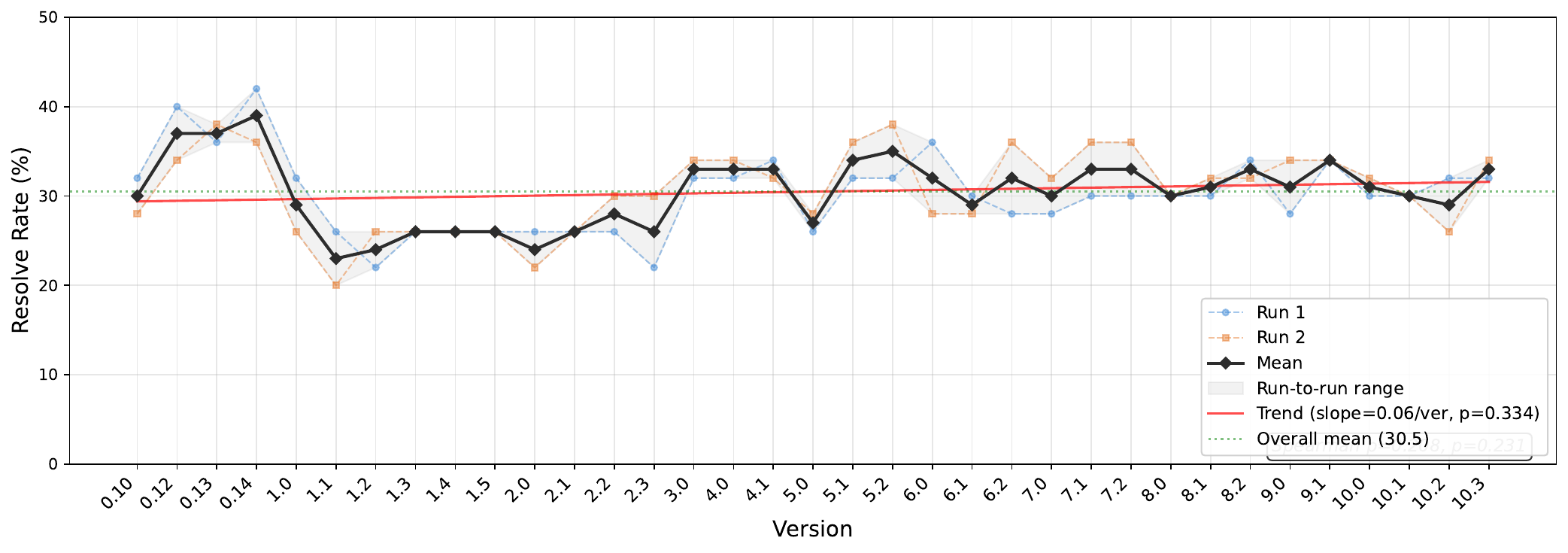}
    \caption{Task Resolve Rate (\%)}
    \label{fig:rq1-raw-resolve}
\end{subfigure}

\vspace{0.5em}
\begin{subfigure}[b]{\textwidth}
    \centering
    \includegraphics[width=\textwidth]{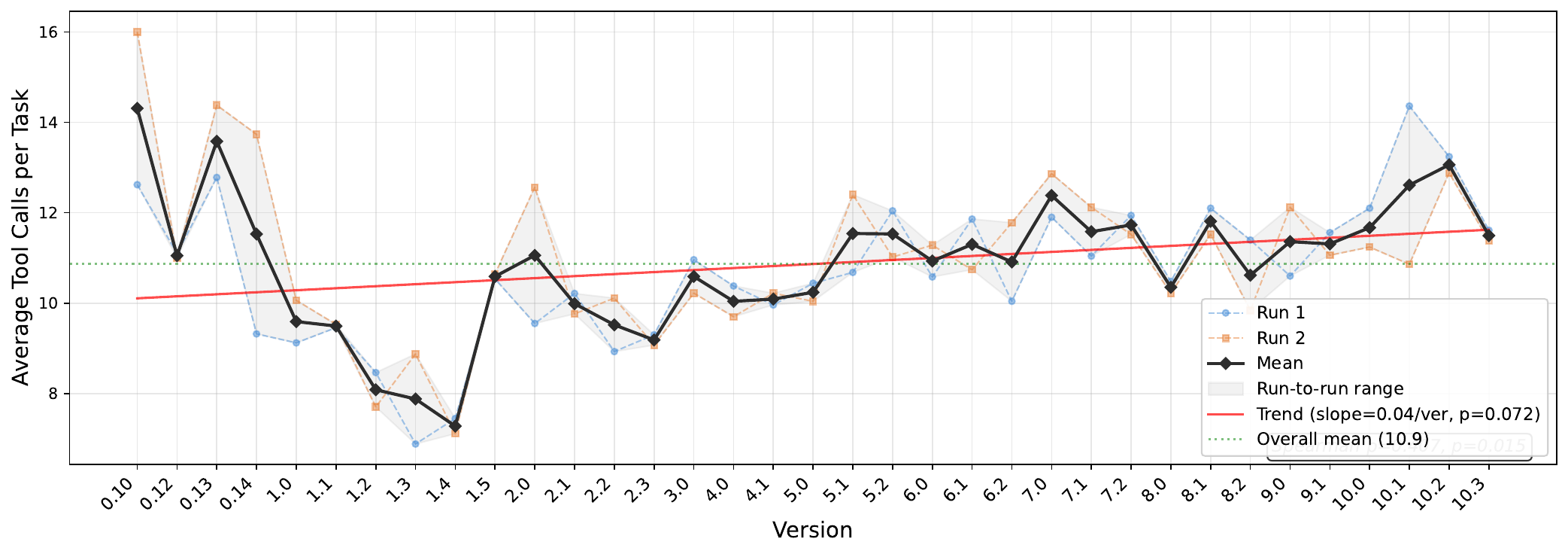}
    \caption{Average Number of Tool Calls per Task}
    \label{fig:rq1-raw-tools}
\end{subfigure}

\vspace{0.5em}
\begin{subfigure}[b]{\textwidth}
    \centering
    \includegraphics[width=\textwidth]{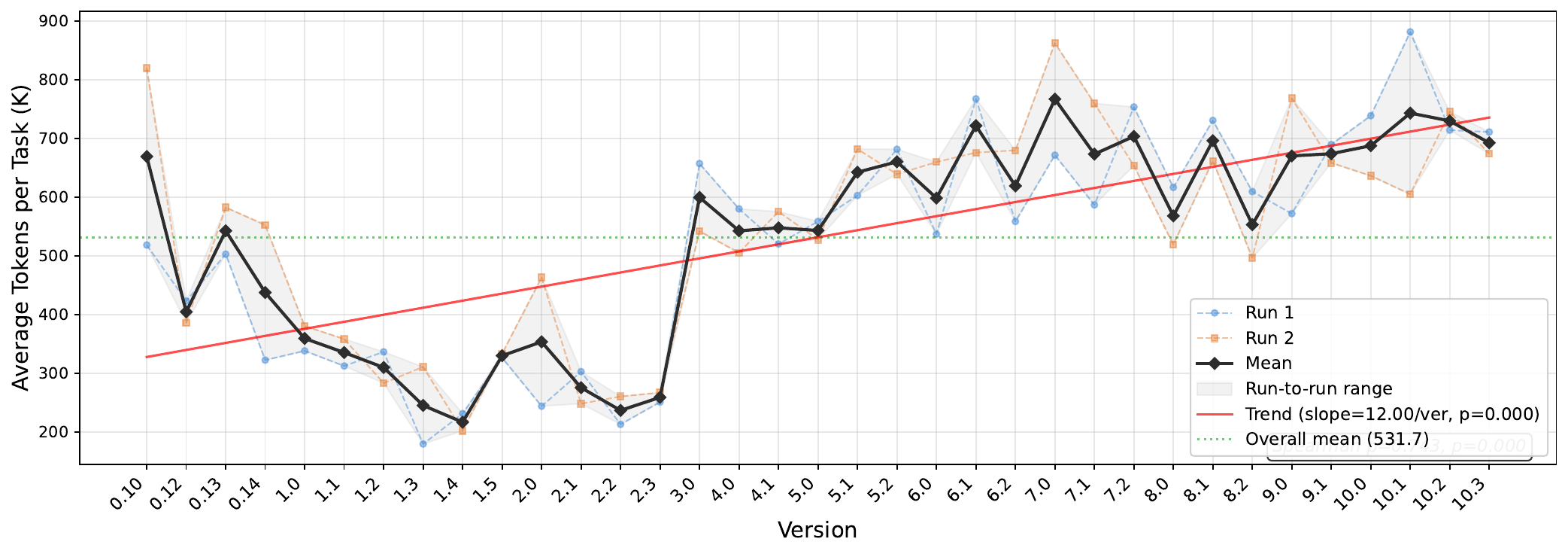}
    \caption{Average API Tokens per Task}
    \label{fig:rq1-raw-tokens}
\end{subfigure}
\caption{Raw quality metrics across 35 sequential releases. The near-perfect overlap between Run~1 and Run~2 confirms high execution stability, ruling out LLM stochasticity as a confound. Version labels in these figures omit the leading '0.' prefix for readability (e.g., '0.14' refers to v0.0.14). Over time, effectiveness remains flat while resource consumption increases.}
\label{fig:rq1-raw}
\end{figure*}

\paragraph{Finding 4: Quality metrics are consistent across independent runs.} Before examining cross-version trends, we established that executing the exact same tasks twice yields consistent results. 
Figures~\ref{fig:rq1-raw-resolve}, \ref{fig:rq1-raw-tokens}, and \ref{fig:rq1-raw-tools} display both runs alongside their mean for resolve rate, token consumption, and tool calls respectively, showing closely aligned Run1 and Run~2 trends across all 35 versions. 
While minor variance exists for individual releases, Wilcoxon signed-rank tests confirm that the overall statistical distributions are identical for resolve rate ($p=0.478$), token consumption ($p=0.432$), and tool calls ($p=0.466$). Furthermore, 87.7\% of task executions demonstrate perfect binary agreement (i.e., resolved in both runs or unresolved in both). Establishing this baseline consistency is critical, as it guarantees that the evolutionary fluctuations explored in the following findings stem from actual changes rather than stochastic LLM noise.

\paragraph{Finding 5: Resolve rate fluctuates without statistically significant trend across releases.}
Despite rapid release cycles, thousands of new commits, and an advancing version number, the resolve rate exhibits substantial fluctuations between releases without a statistically significant monotonic trend. The Spearman rank correlation between chronological release order (i.e., release sequence number, from 1 to 35) and resolve rate is weak and non-significant ($\rho = 0.208$, $p = 0.231$). The resolve rate fluctuates around a stable mean of 30.5\%, but successive releases often differ markedly in quality, particularly during the early evolution of the agent harness (up to approximately v0.3.0). After this period, quality stabilizes into a narrower plateau, although isolated regressions still occur (e.g., around v0.5.0). Quality dipped as low as 23.0\% in some mid-cycle updates, yet peaked at 39.0\% in some of the earliest versions (e.g., v0.0.14), demonstrating that newer releases do not consistently outperform earlier ones.

\begin{figure*}[!htbp]
\centering
\includegraphics[width=\textwidth]{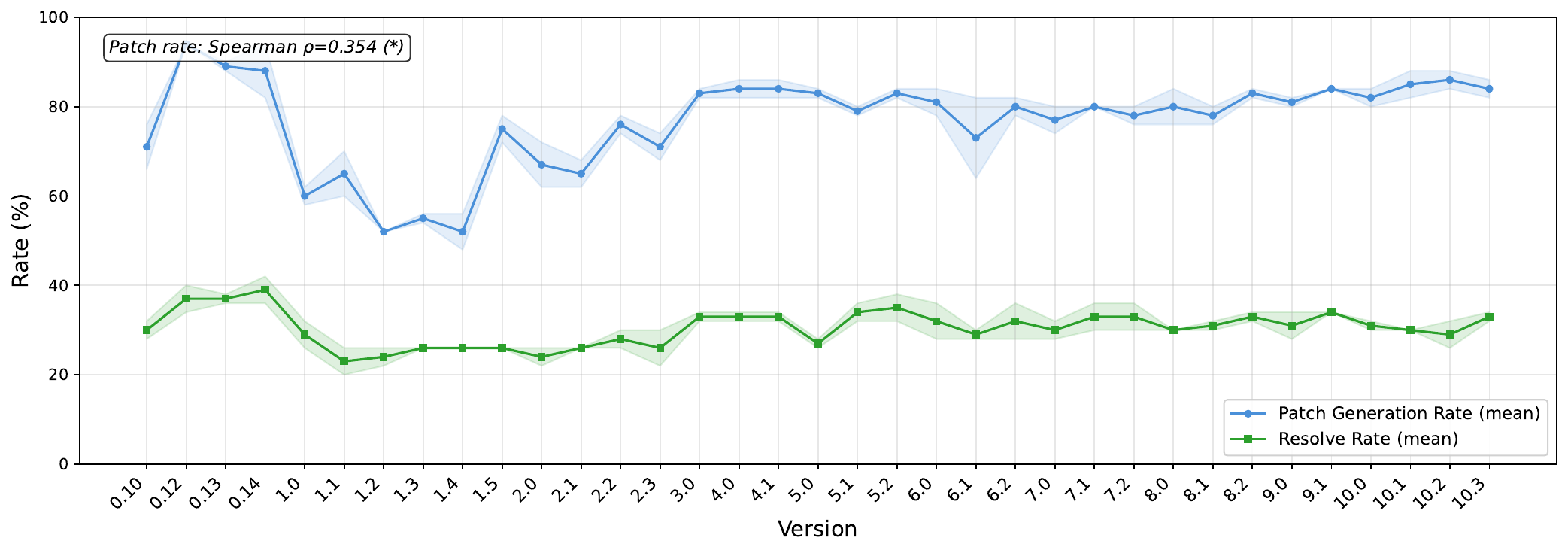}
\caption{Patch generation rate vs.\ resolve rate across 35 releases. Version labels in these figures omit the leading '0.' prefix for readability (e.g., '0.14' refers to v0.0.14). While the agent produces a non-empty patch in 52\%--94\% of tasks, only 23\%--39\% pass the test suite, illustrating a persistent gap between attempting a fix and producing a correct one.}
\label{fig:rq1-patch-vs-resolve}
\end{figure*}

Figure~\ref{fig:rq1-patch-vs-resolve} also shows a disconnect between generating a patch and generating a \textit{correct} patch. Depending on the version, the agent produces a code patch in 52\%--94\% of tasks, but only 23\%--39\% of tasks are resolved correctly. Agent harness updates substantially alter the agent's reasoning behavior, yet these modifications do not push the ceiling of correctness.

\paragraph{Finding 6: Token consumption and number of tool calls exhibit an overall increasing trend with significant variations at intermediate versions.}
Resource consumption followed a markedly different trajectory from resolve rate, growing significantly over the study period, as shown in Figures~\ref{fig:rq1-raw-tools} and \ref{fig:rq1-raw-tokens}. Token consumption exhibits a strong and significant increasing trend over time (Spearman $\rho = 0.743$, $p<0.0001$). Despite a sharp mid-cycle drop to a trough of \textasciitilde217K tokens at v0.1.4, consumption increased again in the latest versions. The first nine releases averaged approximately 391K tokens per task, whereas the latest versions consumed nearly 668K per task, an increase of over 70\% in just a few months with no corresponding improvement in resolve rate.
The average number of tool calls followed the same pattern (Figure~\ref{fig:rq1-raw-tools}), fluctuating between 6.9 and 14.3 across releases.

\begin{figure*}[!htbp]
\centering
\begin{subfigure}[b]{\textwidth}
    \centering
    \includegraphics[width=\textwidth]{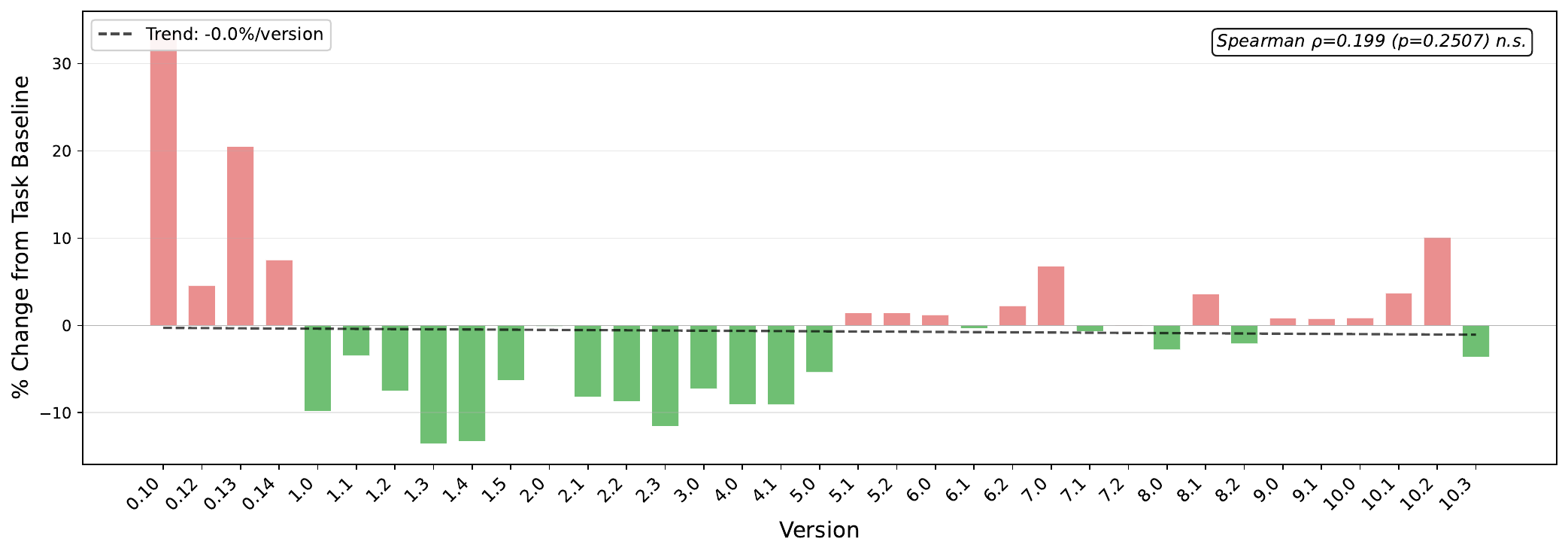}
    \caption{Normalized Tool Calls (\% Change per Task)}
    \label{fig:rq1-norm-tools}
\end{subfigure}

\vspace{0.5em}
\begin{subfigure}[b]{\textwidth}
    \centering
    \includegraphics[width=\textwidth]{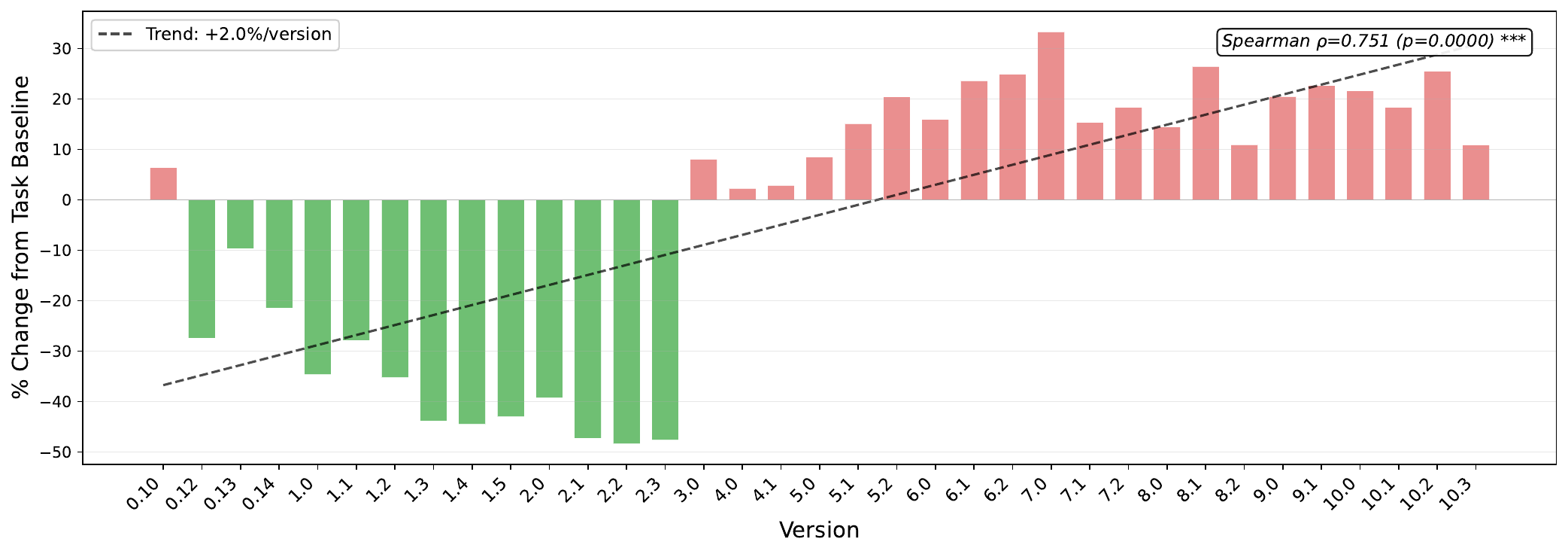}
    \caption{Normalized Token Consumption (\% Change per Task)}
    \label{fig:rq1-norm-tokens}
\end{subfigure}
\caption{Task-normalized quality metrics. Version labels in these figures omit the leading '0.' prefix for readability (e.g., '0.14' refers to v0.0.14). Bars above the zero line (red) indicate an increase relative to that task's baseline (i.e., its mean cost across all 35 versions and both runs). Bars below (green) indicate a decrease. Normalizing relative to per-task baselines confirms that the upward trend in tool calls and token consumption is an agent harness-level phenomenon, not an artifact of hard-task outliers.}
\label{fig:rq1-normalized}
\end{figure*}

As shown in Figure~\ref{fig:rq1-normalized}, the task-normalized analysis described in Section~\ref{subsubsec:methodology-metrics} (where each version's quality value is expressed as a percentage deviation from that task's mean across all versions) confirmed the raw (unnormalized) finding from Figure~\ref{fig:rq1-raw-tokens} for token consumption over time ($\rho = 0.751$, $p<0.0001$). The latest releases consumed up to 19.6\% more tokens per task relative to the per-task baseline, whereas earlier versions operated at 24.2\% below the baseline. Critically, this cost inflation is not offset by any gain in effectiveness as resolve rate peaked at \textasciitilde39\% early and never recovered, meaning that later versions spend significantly more resources to achieve the same or worse task resolution.

\paragraph{Finding 7: Failed tasks consume more resources than successful ones.}
There is no significant positive correlation between token usage and task success across versions ($\rho \approx -0.02$, $p=0.91$). On average, successfully resolved tasks require 7.2 tool calls and 258.7K tokens, compared to 12.95 tool calls and 697.7K tokens for unresolved ones (Figures~\ref{fig:rq1-raw-tools} and~\ref{fig:rq1-raw-tokens}), which means unresolved tasks consume nearly 2.7x more tokens and invoke 1.8x more tool calls. A plausible explanation is that correct fixes are recognized and executed quickly, whereas failing tasks trap the model in fruitless edit-test-read loops that consume tokens without improving the generated patches.

\begin{figure}[htbp]
\centering
\includegraphics[width=\columnwidth]{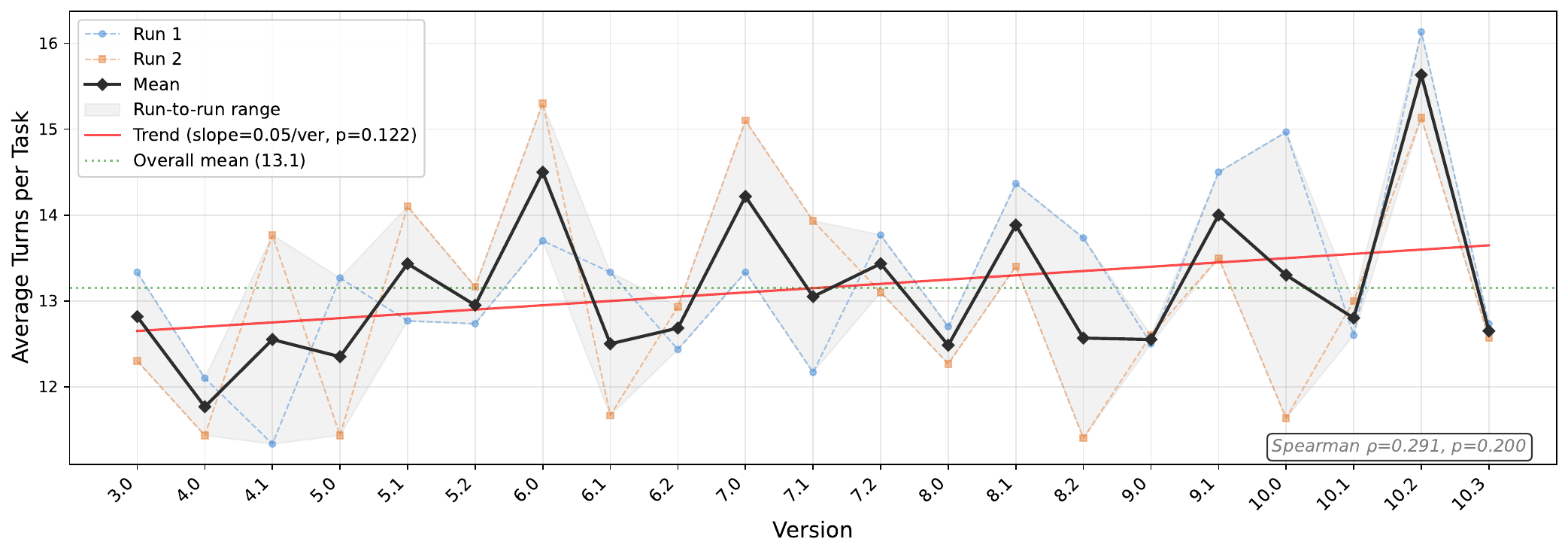}
\caption{Evolution of the number of conversation turns across versions. Later versions drive more LLM exchanges, leading to increased context accumulation.}
\label{fig:rq1-turns}
\end{figure}

\paragraph{Finding 8: Larger system prompts and more conversation turns compound to drive token inflation.}
Trajectory analysis (i.e., inspecting the structured execution logs for individual task runs, including the initial prompt payload and the sequence of LLM turns and tool calls) suggests a possible explanation for this token inflation as two architectural changes compound to drive this cost increase. First, the initial prompt payload at step 1 (comprising the system prompt, tool schemas, and task description) grew by \textasciitilde8\% from the earliest to the latest releases (averaged over the first and last five versions), with the task description held constant across versions. This means that this payload growth is attributable primarily to agent harness changes, specifically the expansion of system prompt instructions and tool schemas. Second, newer releases require 18\% more LLM turns on average (Figure~\ref{fig:rq1-turns}). This is confirmed by a near-perfect correlation between token consumption and LLM turns ($\rho=0.941$, $p<0.0001$). Because agent architectures prepend the full conversation history at each API call, more turns mechanically inflate token consumption. Since the larger system prompt is included at every turn, its overhead is repeatedly incurred, causing the initial prompt expansion to compound across the increased number of turns.

\subsection{Case Studies}

To better understand the quantitative results of this RQ, we manually examined two representative cases to trace specific code changes behind the quantitative trends observed above.

The transition from v0.1.4 to v0.1.5 (Figure~\ref{fig:rq1-raw-tokens}) illustrates how agent harness changes can dramatically inflate resource consumption without any gain in effectiveness. v0.1.4 was the most token-efficient release, consuming 216.6K tokens and 7.28 tool calls per task at a 26.0\% resolve rate. One day later, v0.1.5 was released with the same resolve rate (26.0\%), but token consumption surged 52\% to 329.9K, and tool calls jumped to 10.59 per task. Of the six PRs merged in this release, two directly affected agent behavior and drove this inflation.
PR~\#969\footnote{\url{https://github.com/QwenLM/qwen-code/pull/969}} (``Simplify and Improve Search Tools'') rewrote the glob, grep, and ripgrep tool implementations (+834/$-$776 lines), fundamentally changing how the agent explores codebases. PR~\#981\footnote{\url{https://github.com/QwenLM/qwen-code/pull/981}} (``Customizable Model Training and Tool Output Management'') restructured how tool outputs are formatted and presented back to the model (+795/$-$607 lines). Each of these changes passed all unit and integration tests in the project's CI pipeline, yet no automated check flagged the 52\% increase in token consumption.

The transition from v0.2.3 to v0.3.0 (Figure~\ref{fig:rq1-raw-tokens}) saw token consumption more than double, from 259K to 599.6K per task (+131\%), with the number of tool calls rising from 9.19 to 10.59, while the resolve rate improved from 26.0\% to 33.0\%. This release window included PR~\#926\footnote{\url{https://github.com/QwenLM/qwen-code/pull/926}} (``Headless enhancement: add \texttt{stream-json} as input/output format''), the single largest change in our study at +14,550 lines, which restructured the CLI's streaming I/O layer, and PR~\#1058\footnote{\url{https://github.com/QwenLM/qwen-code/pull/1058}} (``Add Internationalization Support for UI and LLM Output''), which injected +4,740 lines of i18n infrastructure into the system prompt and UI paths. Simultaneously, PR~\#1076\footnote{\url{https://github.com/QwenLM/qwen-code/pull/1076}} removed the prompt completion feature ($-$671 lines). Again, every PR passed the project's automated checks. No automated check flagged the 131\% increase in token consumption or the architectural restructuring of the streaming layer.

These examples represent the \emph{most visible} development patterns we observe throughout the releases: large-scale agent harness changes that pass all automated checks while silently inflating resource consumption. In the absence of regression testing that evaluates agentic quality (resolve rate, token efficiency, tool call overhead), code changes that fundamentally alter agent behavior are merged based solely on whether the underlying agent executes without error.


\begin{findingbox}[Summary of RQ1]
Chronological iteration of agent harness releases does not guarantee improved quality. Later agent harness versions consume nearly double the tokens, driven by 8\% larger base prompts and 18\% more reasoning turns, while resolve rates peaked early at \textasciitilde39\% and never recovered their peak. Unresolved tasks consume 2.7$\times$ more tokens and 1.8$\times$ more tool calls than resolved ones, challenging the assumption that heavier agent harness yields stronger autonomous quality.
\end{findingbox}
\section{RQ2: What Project-Level Development Patterns Explain Quality Shifts?}
\label{sec:rq2}

\subsection{Motivation}

Knowing that quality shifts occur across releases is not sufficient for practitioners to act on them; we need to understand which development patterns drive them. Prior work has shown that release characteristics such as code churn, commit composition, and contributor activity are associated with software quality outcomes in traditional software projects~\cite{nagappan2005churn, khomh2015rapid}. However, it is unclear whether these relationships hold for agent harness. We therefore examine whether certain release patterns, such as feature-heavy vs. fix-heavy releases, large vs. small changes, and release cadence, are associated with quality improvements or regressions, to help developers make more informed decisions about how they ship agent harness changes.

\subsection{Study Design}
\label{sec:rq2:design}

In RQ1, we tracked quality across three dimensions: Resolve Rate (the percentage of tasks the agent successfully solves), Token Consumption (the average number of LLM tokens consumed per task), and Tool Calls (the average number of tool invocations per task).


In order to understand potential factors associated with quality shifts, we extract 22 release-level factors for each release window, enumerated in Table~\ref{tab:rq2-factors}. The raw data (e.g., commits, pull requests, issue reports, and code churn) is collected via the GitHub API. Factors such as \texttt{feat\_churn\_ratio} and \texttt{fix\_pr\_ratio} are derived by classifying each PR using the conventional commits specification\footnote{\url{https://www.conventionalcommits.org/}}. These factors are grounded in release engineering and empirical SE literature~\cite{nagappan2005churn, nagappan2008influence, kerzazi2014factors, gousios2014pullbased, adams2016release, hindle2008large}, covering change size, release pace, commit composition, and issue activity.

\begin{table*}[t]
\centering
\caption{The 22 release-level factors extracted across all 35 evaluated versions.}
\label{tab:rq2-factors}
\begin{tabular}{r l l}
\toprule
\textbf{\#} & \textbf{Factor} & \textbf{Description} \\
\midrule
\multicolumn{3}{c}{\textit{Release Cadence}~\cite{adams2016release}} \\
\midrule
1 & \texttt{days\_since\_prev} & \# Days elapsed between this release and the previous one \\
\midrule
\multicolumn{3}{c}{\textit{Development Activity}} \\
\midrule
2 & \texttt{total\_commits} & \# Commits merged in the release window \\
3 & \texttt{pr\_count} & \# Pull requests merged \\
4 & \texttt{unique\_authors} & \# Distinct contributing developers \\
\midrule
\multicolumn{3}{c}{\textit{Code Churn}~\cite{nagappan2005churn}} \\
\midrule
5 & \texttt{total\_churn} & \#Total lines added + deleted across all PRs \\
6 & \texttt{total\_additions} & Total \#lines added \\
7 & \texttt{total\_deletions} & Total \#lines deleted \\
8 & \texttt{mean\_pr\_size} & Average PR size in changed lines (\texttt{total\_churn} / \texttt{pr\_count}) \\
9 & \texttt{max\_pr\_size} & Largest single PR in terms of lines changed \\
10 & \texttt{total\_changed\_files} & \# Files touched \\
\midrule
\multicolumn{3}{c}{\textit{Composition}~\cite{hindle2008large}} \\
\midrule
11 & \texttt{feat\_churn\_ratio} & Fraction of churn from feature-oriented PRs \\
12 & \texttt{fix\_churn\_ratio} & Fraction of churn from bug-fix PRs \\
13 & \texttt{refactor\_churn\_ratio} & Fraction of churn from refactoring PRs \\
14 & \texttt{fix\_pr\_ratio} & Proportion of PRs categorized as fixes \\
15 & \texttt{feat\_pr\_ratio} & Proportion of PRs categorized as features \\
16 & \texttt{fix\_to\_feat\_ratio} & Ratio of maintenance effort to new feature development \\
\midrule
\multicolumn{3}{c}{\textit{Process Quality}~\cite{gousios2014pullbased}} \\
\midrule
17 & \texttt{mean\_time\_to\_merge} & Average hours from PR creation to merge \\
\midrule
\multicolumn{3}{c}{\textit{Project Health}} \\
\midrule
18 & \texttt{needs\_triage\_count} & \# Open issues tagged \texttt{needs-triage} at release time \\
19 & \texttt{issues\_opened} & \# Issues opened during the release window \\
20 & \texttt{bug\_count} & \# Issues labeled as bugs \\
21 & \texttt{feature\_count} & \# Issues categorized as feature requests \\
22 & \texttt{bug\_to\_feature\_ratio} & Ratio of \#bug reports to \#feature requests \\
\bottomrule
\end{tabular}
\end{table*}

\paragraph{Classification Thresholding.}

\begin{figure}[htbp]
\centering
\includegraphics[width=\columnwidth]{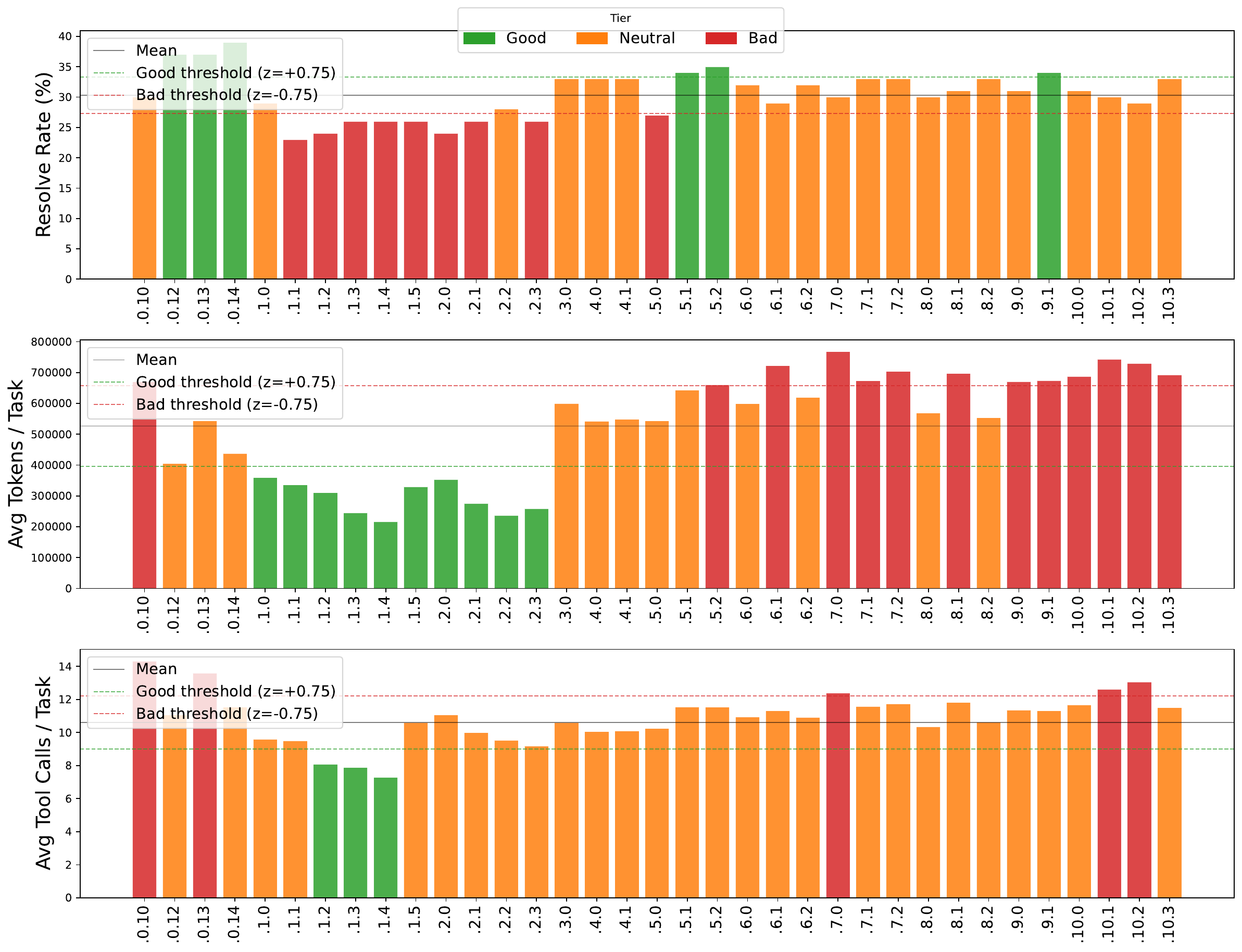}
\caption{Distribution of quality tiers per dimension ($z = \pm0.75\sigma$ thresholds). Each bar represents one of the 35 evaluated Qwen Code releases (x-axis: version number, with the leading \texttt{v0.} prefix omitted for readability), colored by tier: Good (green), Neutral (orange), or Bad (red). Dashed lines show the Good and Bad classification thresholds.}
\label{fig:rq2-classification}
\end{figure}

We classify the 35 releases of \qwen~harness into three tiers based on their $z$-score relative to the mean, i.e., Good, Neutral, and Bad (Figure~\ref{fig:rq2-classification}), to distinguish the highest and lowest-performing versions for later analysis. We use a threshold of $z = \pm0.75\sigma$, which places roughly the top and bottom quarter of releases in each extreme tier while keeping enough versions in each group for meaningful comparison~\cite{cliff1993dominance}.

\begin{itemize}
    \item \textbf{Good Releases ($z \ge +0.75\sigma$):} Versions performing above the mean by at least three-quarters of a standard deviation. For Resolve Rate, this means solving more tasks; for Token Consumption and Tool Calls, it means consuming fewer resources.
    \item \textbf{Bad Releases ($z \le -0.75\sigma$):} Underperforming versions. Note that classifications are computed independently per dimension; a version can be ``Good'' on Resolve Rate while ``Bad'' on Token Consumption.
    \item \textbf{Neutral Releases:} Versions clustering near the mean.
\end{itemize}

\paragraph{Statistical Methods.}

To identify which of the 22 factors in Table~\ref{tab:rq2-factors} are associated with quality shifts, we employ two complementary statistical approaches that capture different aspects of this relationship:

\begin{itemize}
    \item \textbf{Spearman Rank Correlation (Continuous Trends):} Measures the strength and direction of monotonic relationships between a factor and a quality metric across all 35 releases~\cite{spearman1904proof}. The accompanying $p$-value tests whether the correlation is significantly different from zero.
    \item \textbf{Mann-Whitney U \& Cliff's Delta (Extreme Separation):} Compares the distribution of each release-level factor between Good and Bad tiers directly. The Mann-Whitney U test~\cite{mann1947test} evaluates statistical significance, while Cliff's $d$~\cite{cliff1993dominance, vargha2000critique} quantifies the effect size ($|d| \ge 0.474$ indicates a large effect).
\end{itemize}

Together, Spearman captures whether a factor is correlated with quality globally across all releases, while Mann-Whitney and Cliff's $d$ capture whether that factor distinguishes the best from the worst performing releases specifically.

\paragraph{Multiple Testing Correction.}
Given the number of factors evaluated across multiple quality dimensions, uncorrected significance thresholds would inflate the false discovery rate. All $p$-values reported in RQ2 are adjusted using the Benjamini-Hochberg (BH) procedure~\cite{benjamini1995controlling} to control the false discovery rate at $\alpha = 0.05$. Only factors with adjusted $p$-values below this level are reported as significant.

\paragraph{Absolute vs.\ Delta Analysis.}

Absolute metrics (\eg total \texttt{bug\_count}) capture cumulative long-term trends but conflate agent harness effects with project maturity, since later releases naturally accumulate more changes, contributors, and reported issues regardless of harness quality.
Delta ($\Delta$) metrics, by contrast, measure the change from the immediately preceding version, isolating the effect of a single release from the background maturity trend. The two analyses are complementary. The absolute analysis reveals which dimensions of agent harness development consistently associate with quality over the full history, while the delta analysis identifies which release-to-release transitions produce immediate regressions or gains. Together they allow us to distinguish sustained structural patterns from short-term release effects.


\subsection{Results}
\label{sec:rq2:results}

\subsubsection{Absolute Analysis: Accumulated Release Characteristics}
\label{sec:rq2:absolute}
We first assess the accumulated release characteristics across the full timeline. The complete set of absolute statistical results is detailed in Table~\ref{tab:rq2-raw-stats}.

\begin{table*}[t]
\centering

\caption{Absolute (Raw) statistical results. 
$\rho$~=~Spearman rank correlation; 
$p_\rho$~=~$p$-value for the Spearman test; 
$d$~=~Cliff's $\delta$ effect size (Mann-Whitney U comparison between Good and Bad tiers); 
$p_\text{MW}$~=~BH-adjusted $p$-value for the Mann-Whitney U test. 
Significant results ($p<0.05$) are \textbf{bold}; 
$d$ is omitted for non-significant Mann-Whitney results.}

\label{tab:rq2-raw-stats}
\resizebox{\textwidth}{!}{
\begin{tabular}{l cccc cccc cccc}
\toprule
 & \multicolumn{4}{c}{Resolve Rate} & \multicolumn{4}{c}{Token Consumption} & \multicolumn{4}{c}{\#Tool Calls} \\
\cmidrule(lr){2-5} \cmidrule(lr){6-9} \cmidrule(lr){10-13}
Factor & $\rho$ & $p_\rho$ & $d$ & $p_\text{MW}$ & $\rho$ & $p_\rho$ & $d$ & $p_\text{MW}$ & $\rho$ & $p_\rho$ & $d$ & $p_\text{MW}$ \\
\midrule
\texttt{days\_since\_prev}           & 0.230  & 0.1830 & ---    & 0.1471 & -0.035 & 0.8442 & ---    & 0.9264 & -0.157 & 0.3682 & ---    & 1.0000 \\
\texttt{total\_commits}              & 0.233  & 0.1716 & ---    & 0.6250 &  0.417 & \textbf{0.0114} & -0.601 & \textbf{0.0138} &  0.140 & 0.4166 & ---    & 0.3893 \\
\texttt{pr\_count}                   & 0.122  & 0.4778 & ---    & 0.2767 &  0.252 & 0.1387 & ---    & 0.3840 &  0.118 & 0.4913 & ---    & 0.2187 \\
\texttt{unique\_authors}             & 0.060  & 0.7279 & ---    & 0.9120 &  0.428 & \textbf{0.0091} & -0.482 & \textbf{0.0476} &  0.193 & 0.2600 & ---    & 0.3252 \\
\texttt{total\_churn}                & 0.216  & 0.2056 & ---    & 0.3676 &  0.065 & 0.7048 & ---    & 0.6851 & -0.054 & 0.7531 & ---    & 0.7302 \\
\texttt{total\_additions}            & 0.246  & 0.1485 & ---    & 0.2198 &  0.141 & 0.4127 & ---    & 0.3539 & -0.002 & 0.9917 & ---    & 0.5556 \\
\texttt{total\_deletions}            & 0.142  & 0.4081 & ---    & 0.4278 & -0.083 & 0.6313 & ---    & 0.5820 & -0.096 & 0.5780 & ---    & 0.8057 \\
\texttt{mean\_pr\_size}              & 0.219  & 0.1987 & ---    & 0.4278 & -0.082 & 0.6362 & ---    & 0.7281 & -0.161 & 0.3470 & ---    & 0.9048 \\
\texttt{max\_pr\_size}               & 0.227  & 0.1824 & ---    & 0.3676 & -0.018 & 0.9169 & ---    & 1.0000 & -0.102 & 0.5521 & ---    & 0.9048 \\
\texttt{total\_changed\_files}       & 0.232  & 0.1732 & ---    & 0.1927 &  0.125 & 0.4677 & ---    & 0.6223 &  0.027 & 0.8745 & ---    & 0.1905 \\
\texttt{feat\_churn\_ratio}          & 0.438  & \textbf{0.0075} & ---    & 0.0793 &  0.401 & \textbf{0.0154} & -0.573 & \textbf{0.0149} &  0.360 & \textbf{0.0309} & -0.900 & \textbf{0.0365} \\
\texttt{fix\_churn\_ratio}           & 0.112  & 0.5157 & ---    & 0.2992 &  0.367 & \textbf{0.0276} & -0.496 & \textbf{0.0421} &  0.297 & 0.0781 & ---    & 0.2857 \\
\texttt{refactor\_churn\_ratio}      & -0.372 & \textbf{0.0254} & ---    & 0.1013 & -0.196 & 0.2517 & ---    & 0.2089 & -0.214 & 0.2098 & ---    & 0.8668 \\
\texttt{fix\_pr\_ratio}              & 0.166  & 0.3336 & ---    & 0.2989 &  0.356 & \textbf{0.0332} & ---    & 0.1034 &  0.166 & 0.3345 & ---    & 0.3252 \\
\texttt{feat\_pr\_ratio}             & 0.351  & \textbf{0.0358} & ---    & 0.1404 &  0.334 & \textbf{0.0462} & -0.489 & \textbf{0.0379} &  0.265 & 0.1190 & ---    & 0.1761 \\
\texttt{fix\_to\_feat\_ratio}        & 0.048  & 0.7820 & ---    & 0.5076 &  0.165 & 0.3359 & ---    & 0.6600 &  0.019 & 0.9114 & ---    & 0.8041 \\
\texttt{mean\_time\_to\_merge\_hours}& 0.185  & 0.2791 & ---    & 0.1806 &  0.108 & 0.5322 & ---    & 0.9078 & -0.080 & 0.6436 & ---    & 0.9048 \\
\texttt{needs\_triage\_count}        & 0.330  & \textbf{0.0496} & ---    & 0.4150 &  0.566 & \textbf{0.0003} & -0.832 & \textbf{0.0005} &  0.308 & 0.0676 & ---    & 0.0733 \\
\texttt{issues\_opened}              & 0.168  & 0.3264 & ---    & 0.1924 & -0.068 & 0.6935 & ---    & 0.7058 & -0.064 & 0.7130 & ---    & 0.6228 \\
\texttt{bug\_count}                  & 0.463  & \textbf{0.0045} & 0.800  & \textbf{0.0096} &  0.319 & 0.0578 & -0.545 & \textbf{0.0243} &  0.283 & 0.0938 & ---    & 0.1706 \\
\texttt{feature\_count}              & 0.331  & \textbf{0.0483} & 0.617  & \textbf{0.0311} &  0.320 & 0.0574 & ---    & 0.0509 &  0.295 & 0.0808 & ---    & 0.3812 \\
\texttt{bug\_to\_feature\_ratio}     & 0.546  & \textbf{0.0006} & 0.750  & \textbf{0.0061} &  0.362 & \textbf{0.0302} & -0.664 & \textbf{0.0036} &  0.367 & \textbf{0.0276} & ---    & 0.1248 \\
\bottomrule
\end{tabular}}
\end{table*}

\begin{figure}[htbp]
\centering
\includegraphics[width=\columnwidth]{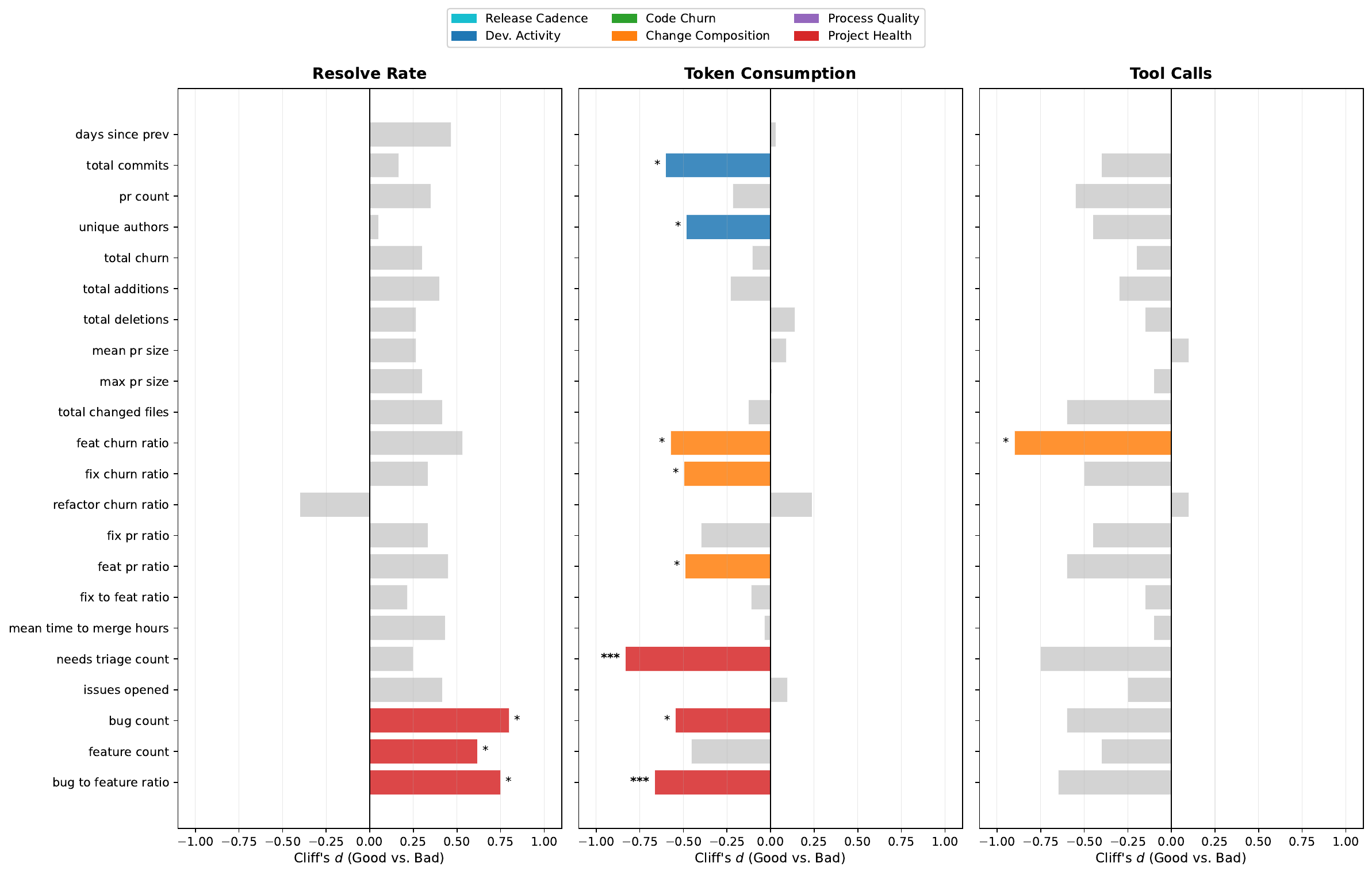}
\caption{Cliff's delta effect sizes comparing Good vs. Bad tiers across release-level factors. Factors are colored only when statistically significant ($p < 0.05$). Non-significant factors are shown in grey.}
\label{fig:rq2-forest}
\end{figure}


\paragraph{Finding 9: Development intensity correlates with cost, but not with improved effectiveness.}
Releases with more commits and contributing authors systematically correlate with increased token consumption without improving agent ability to solve tasks (Table~\ref{tab:rq2-raw-stats}). 
For instance, \texttt{total\_commits} ($d=-0.601$) and \texttt{unique\_authors} ($d=-0.482$) both show large negative effect sizes for token consumption. This indicates that bad releases (i.e., costly releases) have systematically more commits and more contributors than cheap ones, as shown in Figure~\ref{fig:rq2-forest}.
One plausible explanation is that more commits and more contributors tend to introduce additional tools, configuration, and prompt changes that accumulate over the release, increasing the LLM's processing burden without improving task resolution.

\paragraph{Finding 10: Feature and issue activity are positively associated with resolve rate, but at the cost of efficiency.}
Several factors show significant positive associations with the resolve rate. Issue-related factors show the strongest signals in terms of effect size: \texttt{bug\_to\_feature\_ratio} ($\rho=0.546$, $p=0.0006$) and \texttt{bug\_count} ($\rho=0.463$, $p=0.0045$) are the strongest correlates overall, while \texttt{feature\_count} ($\rho=0.331$, $p=0.0483$) also reaches significance. Among composition factors, \texttt{feat\_churn\_ratio} ($\rho=0.438$, $p=0.0075$) and \texttt{feat\_pr\_ratio} ($\rho=0.351$, $p=0.0358$) both correlate positively with Resolve Rate. 
However, \texttt{feat\_churn\_ratio} also correlates with higher Token Consumption ($\rho=0.401$, $d=-0.573$) and more Tool Calls ($\rho=0.360$, $d=-0.900$), revealing a trade-off between effectiveness and efficiency in feature-heavy releases. New features typically expand the agent's tool set and system prompt, giving the LLM more capabilities to reason over but also more context to process on every turn.

A concrete example illustrates both the potential and the cost of feature work. The transition from v0.5.0 to v0.5.1 saw resolve rate jump from 27\% to 34\% (Figure~\ref{fig:rq1-raw-resolve}), the largest single-release improvement in our dataset. This release included PR~\#1269 (``fix: default values of sampling params''), a five-line change to sampling parameter defaults that may have improved the model's reasoning quality, and PR~\#1214 (``add configurable OpenAPI 3.0 schema compliance''), which improved tool definition formatting. However, token consumption also rose from 561K to 642K (+14\%), consistent with the feature-efficiency trade-off our statistical analysis identifies.

\paragraph{Finding 11: Refactoring and release cadence show no quality benefit.}
\texttt{refactor\_churn\_ratio} is the only factor negatively associated with Resolve Rate ($\rho=-0.372$, $p=0.025$), with no corresponding efficiency gains, suggesting that refactoring reorganizes internal logic without improving task-solving capability and may temporarily disrupt established reasoning pathways. Neither \texttt{days\_since\_prev} nor \texttt{mean\_pr\_size} is significantly associated with any quality metric, indicating that release timing and PR size alone do not predict agent quality.


\subsubsection{Delta ($\Delta$) Analysis: Immediate Release-to-Release Transitions}
\label{sec:rq2:delta}
To remove longitudinal maturity effects, we analyze the delta between consecutive versions. The complete delta results are presented in Table~\ref{tab:rq2-delta-stats}.

\begin{table*}[t]
\centering
\caption{Delta ($\Delta$) statistical results tracking immediate release transitions. $\rho$~=~Spearman correlation, $d$~=~Cliff's $\delta$, $p_\rho$~=~Spearman $p$-value, $p_\text{MW}$~=~BH-adjusted Mann-Whitney $p$-value). Significant $p$-values ($p<0.05$) are \textbf{bold}; Cliff's $d$ is omitted for non-significant Mann-Whitney results.}
\label{tab:rq2-delta-stats}
\resizebox{\textwidth}{!}{
\begin{tabular}{l cccc cccc cccc}
\toprule
 & \multicolumn{4}{c}{Resolve Rate} & \multicolumn{4}{c}{Token Consumption} & \multicolumn{4}{c}{Tool Calls} \\
\cmidrule(lr){2-5} \cmidrule(lr){6-9} \cmidrule(lr){10-13}
Factor & $\rho$ & $p_\rho$ & $d$ & $p_\text{MW}$ & $\rho$ & $p_\rho$ & $d$ & $p_\text{MW}$ & $\rho$ & $p_\rho$ & $d$ & $p_\text{MW}$ \\
\midrule
\texttt{mean\_time\_to\_merge\_hours} &  0.296 & 0.089 & --- & 0.456 & -0.246 & 0.161 & --- & 0.755 & -0.276 & 0.114 & --- & 0.232 \\
\texttt{mean\_pr\_size}               &  0.258 & 0.141 & --- & 0.805 & -0.441 & \textbf{0.009} & --- & 0.081 & -0.484 & \textbf{0.004} & --- & 0.054 \\
\texttt{unique\_authors}              & -0.249 & 0.155 & --- & 0.054 & -0.021 & 0.907 & --- & 0.698 & -0.082 & 0.644 & --- & 0.600 \\
\texttt{fix\_churn\_ratio}            & -0.243 & 0.165 & --- & 0.383 &  0.339 & 0.050 & -0.833 & \textbf{0.008} &  0.255 & 0.145 & --- & 0.336 \\
\texttt{feat\_pr\_ratio}              &  0.174 & 0.325 & --- & 0.902 & -0.012 & 0.947 & --- & 0.755 & -0.076 & 0.668 & --- & 0.954 \\
\texttt{pr\_count}                    & -0.171 & 0.332 & --- & 0.096 & -0.126 & 0.479 & --- & 1.000 & -0.115 & 0.517 & --- & 0.336 \\
\texttt{feat\_churn\_ratio}           &  0.160 & 0.367 & --- & 0.710 &  0.088 & 0.620 & --- & 0.662 &  0.138 & 0.436 & --- & 0.281 \\
\texttt{total\_commits}               & -0.156 & 0.379 & --- & 0.159 & -0.098 & 0.583 & --- & 0.897 & -0.106 & 0.549 & --- & 0.463 \\
\texttt{bug\_to\_feature\_ratio}      &  0.145 & 0.414 & --- & 0.535 & -0.018 & 0.918 & --- & 0.244 & -0.033 & 0.854 & --- & 0.861 \\
\texttt{refactor\_churn\_ratio}       & -0.129 & 0.467 & --- & 0.104 & -0.276 & 0.114 & --- & 0.259 & -0.070 & 0.695 & --- & 0.267 \\
\texttt{total\_deletions}             & -0.129 & 0.469 & --- & 0.165 & -0.351 & \textbf{0.042} & --- & 0.282 & -0.329 & 0.058 & --- & 0.152 \\
\texttt{fix\_to\_feat\_ratio}         &  0.117 & 0.509 & --- & 0.406 & -0.058 & 0.745 & --- & 0.746 & -0.108 & 0.541 & --- & 0.602 \\
\texttt{feature\_count}               & -0.116 & 0.516 & --- & 0.405 & -0.078 & 0.661 & --- & 0.650 & -0.108 & 0.544 & --- & 0.520 \\
\texttt{bug\_count}                   &  0.068 & 0.701 & --- & 0.749 &  0.005 & 0.977 & --- & 0.698 & -0.039 & 0.825 & --- & 0.562 \\
\texttt{max\_pr\_size}                &  0.043 & 0.810 & --- & 0.535 & -0.290 & 0.096 & --- & 0.414 & -0.351 & \textbf{0.042} & --- & 0.121 \\
\texttt{days\_since\_prev}            &  0.043 & 0.811 & --- & 0.628 & -0.098 & 0.586 & --- & 0.524 & -0.122 & 0.498 & --- & 0.209 \\
\texttt{total\_churn}                 &  0.036 & 0.841 & --- & 0.456 & -0.291 & 0.095 & --- & 0.491 & -0.312 & 0.072 & --- & 0.232 \\
\texttt{needs\_triage\_count}         & -0.031 & 0.861 & --- & 0.367 & -0.025 & 0.887 & --- & 0.603 & -0.044 & 0.804 & --- & 0.637 \\
\texttt{total\_changed\_files}        & -0.024 & 0.894 & --- & 0.383 & -0.267 & 0.127 & --- & 0.491 & -0.258 & 0.141 & --- & 0.336 \\
\texttt{fix\_pr\_ratio}               &  0.018 & 0.920 & --- & 0.902 &  0.084 & 0.636 & --- & 0.282 &  0.007 & 0.968 & --- & 1.000 \\
\texttt{issues\_opened}               & -0.006 & 0.971 & --- & 0.441 & -0.014 & 0.939 & --- & 0.332 & -0.074 & 0.679 & --- & 0.602 \\
\texttt{total\_additions}             & -0.003 & 0.987 & --- & 0.318 & -0.233 & 0.184 & --- & 0.573 & -0.281 & 0.108 & --- & 0.189 \\
\bottomrule
\end{tabular}}
\end{table*}

\paragraph{Finding 12: Larger PR size is associated with immediate cost savings.}
Increases in \texttt{mean\_pr\_size} correlate with decreased Token Consumption ($\rho=-0.441$, $p=0.009$) and fewer Tool Calls ($\rho=-0.484$, $p=0.004$). 
Micro-PRs that address narrow, isolated changes may incrementally add edge-case handling and configuration overhead to the agent harness without providing proportional capability gains, whereas larger PRs tend to deliver more self-contained, coherent changes.

\paragraph{Finding 13: Code cleanup associates with lower token consumption but fix-heavy transitions correlate with higher costs.}
Transitions with higher \texttt{total\_deletions} correspond to reduced Token Consumption ($\rho=-0.351$, $p=0.042$; Table~\ref{tab:rq2-delta-stats}). Removing dead code, unused tool definitions, and deprecated configuration actively trims execution prompts. For example, the transition from v0.4.1 to v0.5.0 removed five entire CLI management modules,\footnote{\url{https://github.com/QwenLM/qwen-code/compare/v0.4.1...v0.5.0}}, eliminating 817 lines of unused agent harness code. Token consumption dropped from 517K to 450K ($-$12.9\%) and tool calls from 10.2 to 9.1, consistent with a leaner agent harness footprint.

The opposite pattern holds for fix-heavy transitions. When the proportion of \texttt{fix\_churn\_ratio} increases relative to the previous release, Token Consumption rises ($\rho=0.339$, $p=0.050$, Cliff's $d=-0.833$). This is the largest effect size observed in the delta analysis. Releases dominated by bug fixes often introduce expensive edge-case handling that adds to the agent's prompt burden without improving resolution capability.

This pattern is concretely illustrated by the transition from v0.1.1 to v0.1.2, which was dominated by bug fixes.\footnote{\url{https://github.com/QwenLM/qwen-code/compare/v0.1.1...v0.1.2}} PR~\#898 (``Bug Fixes Release v0.1.1'') bundled multiple fixes (+343/$-$438 lines), while PR~\#935 (``Fix Chat Compression System Instruction and Empty Summary Edge Case'') repaired the context compression pipeline (+749/$-$245 lines). In this case, fixing the chat compression system actually \emph{improved} token efficiency (tokens dropped from 231K to 170K, a 26\% reduction) because the fix eliminated a pathological failure mode where broken compression caused context to accumulate unboundedly. However, resolve rate remained essentially flat. This example reveals an important nuance; while fix-heavy releases \emph{on average} associate with higher costs (the statistical finding), targeted fixes to efficiency-critical components like context compression can be beneficial.





\begin{findingbox}[Summary of RQ2]
The absolute analysis reveals that \texttt{bug\_to\_feature\_ratio} and \texttt{feat\_churn\_ratio} show the strongest positive associations with resolve rate, though feature-heavy releases also associate with higher token consumption and tool calls. Fix-heavy releases and high development intensity are associated with increased resource consumption without resolution gains. The delta analysis yields far fewer significant results. The clearest signals are that consolidating changes into larger PRs and removing obsolete code are associated with immediate cost reductions. The two analyses are complementary: absolute captures sustained structural patterns, while delta isolates the immediate effect of individual transitions.
\end{findingbox}
\section{RQ3: Which Architectural Components Explain Quality Shifts?}
\label{sec:rq3}

\subsection{Motivation}

RQ2 provided project-level explanations, identifying release-level development patterns that are correlated with agent quality. However, project-level factors treat the agent harness as a black box: they capture how the system was built but cannot locate where within the codebase quality shifts originate. To provide actionable guidance to developers, we need to know which specific architectural components are most sensitive to modification. RQ3 addresses this gap by attributing quality shifts to individual architectural layers, enabling developers to make targeted decisions by applying extra scrutiny before touching high-risk layers and proceeding with confidence when modifying low-risk ones.

\subsection{Study Design}
\label{sec:rq3:design}
We map quality fluctuations to specific architectural components by overlaying the quality metrics from RQ1 onto our ten-component reference architecture (detailed in Appendix~\ref{sec:architecture}), focusing on \qwen~as the subject of our longitudinal study. For each release, we extract the set of files modified across all commits in that release window using GitHub API. Each modified file is then mapped to one of the ten reference components using the file path-to-component taxonomy defined in Appendix~\ref{sec:architecture}. This architecture captures the core components common across agents, such as context management, tools layer, and orchestration layer. Files that span multiple components have their churn equally distributed across those components (e.g., \texttt{sharedTokenManager.ts} belongs to both the LLM Provider and Context Management, hence both component receive half the added and deleted lines of that file). This file-to-component mapping enables us to correlate changes in specific architectural zones with the quality shifts identified in RQ1 and characterized in RQ2. 

Not every component is modified in every release, introducing data sparsity that precludes simple binary (modified/unmodified) analysis. For each component, we therefore measure both the raw volume of code changes (\texttt{total\_churn}) and finer-grained factors that capture the nature of the changes: \texttt{feat\_churn}, \texttt{fix\_ratio}, and \texttt{refactor\_churn}, which reflect whether the modifications to a given component are feature additions, bug fixes, or refactoring.

Table~\ref{tab:rq3-factors} lists the nine component-level factors used in RQ3. These are a purposeful adaptation of the release-level factors from Table~\ref{tab:rq2-factors} (RQ2), restricted to metrics meaningful at the granularity of a single architectural component. Release-level factors capturing project-wide properties such as release cadence, contributor count, process quality, and issue tracker state do not apply at the component level and are excluded. The retained factors characterize the nature of changes within each component by their volume, composition by commit type, and churn directionality (i.e., add or delete).

\begin{table}[t]
\centering
\caption{Component-level factors used in RQ3, derived from per-component 
code churn within each release.}
\label{tab:rq3-factors}
\begin{tabular}{r l l}
\toprule
\textbf{\#} & \textbf{Factor} & \textbf{Description} \\
\midrule
\multicolumn{3}{c}{\textit{Churn Directionality}} \\
\midrule
1 & \texttt{add\_del\_ratio}      & Ratio of lines added to lines deleted \\
2 & \texttt{files\_added\_ratio}  & Fraction of changed files that are new additions \\
3 & \texttt{files\_removed\_ratio}& Fraction of changed files that are deletions \\
\midrule
\multicolumn{3}{c}{\textit{Churn Volume by Commit Type}} \\
\midrule
4 & \texttt{feat\_churn}          & \#Lines changed in feature-oriented commits \\
5 & \texttt{fix\_churn}           & \#Lines changed in bug-fix commits \\
6 & \texttt{refactor\_churn}      & \#Lines changed in refactoring commits \\
\midrule
\multicolumn{3}{c}{\textit{Composition by Commit Type}} \\
\midrule
7 & \texttt{feat\_ratio}          & Fraction of commits categorized as features \\
8 & \texttt{fix\_ratio}           & Fraction of commits categorized as bug fixes \\
9 & \texttt{refactor\_ratio}      & Fraction of commits categorized as refactoring \\
\bottomrule
\end{tabular}
\end{table}

A key challenge is that standard correlations between component-level churn and quality are confounded by overall release size: a large release naturally modifies many components simultaneously, creating spurious correlations. To isolate genuine component-level effects, we compute partial Spearman rank correlations~\cite{conover1999practical} ($\rho_{partial}$) controlling for total codebase churn. A partial correlation measures the association between two variables after removing the linear effect of a confounding variable~\cite{conover1999practical}: concretely, we regress both the component-level factor and the quality metric onto total codebase churn, then correlate the residuals. This removes the inflation effect of large releases, ensuring that observed correlations reflect the specific component's contribution rather than the release's overall magnitude. Signals whose partial correlations collapse relative to their raw values are rejected as confounds.

As in RQ2, we report both an Absolute Analysis (cumulative metrics across the 35 versions) and a Delta Analysis (release-to-release transitions). While the absolute analysis captures long-term trends, it inherits the accumulated state from prior releases. The delta analysis isolates the immediate impact of each release by measuring transitions between consecutive versions. Where both analyses agree, the evidence is strongest. Where only the absolute analysis shows significance, the effect likely reflects gradual long-term accumulation rather than a single architectural change. Where only the delta analysis shows significance, the effect is short-lived and may not persist.

\paragraph{Multiple Testing Correction.}
Evaluating multiple architectural components against multiple quality dimensions creates a substantial risk of false discoveries. All $p$-values in RQ3 are adjusted using the Benjamini-Hochberg (BH) correction~\cite{benjamini1995controlling}, consistent with RQ2. We distinguish between significant results ($p < 0.05$).

\subsection{Results}
\label{sec:rq3:results}

\subsubsection{Absolute Analysis: Accumulated Architectural Impact}
The absolute analysis examines how cumulative code changes in each architectural component associate with overall quality across the 35 evaluated versions. Table~\ref{tab:rq3-ext-abs} presents the full results, including both raw and partial correlations.

\begin{table}[t]
\centering

\caption{Component-level absolute analysis: component-level correlations with quality dimensions, reporting only significant raw results ($p < 0.05$). Non-significant combinations are omitted. Row IDs (AE\#) uniquely identify each result for cross-referencing in the text (AE~=~Absolute Effect). $\rho$~=~Spearman rank correlation; Partial~$\rho$~=~partial Spearman correlation controlling for total codebase churn; $p$~=~$p$-value.}

\label{tab:rq3-ext-abs}
\begin{tabular}{c l l l r r r}
\toprule
\# & Dimension & Component & Metric & Spearman $\rho$ & $p$ & Partial $\rho$ \\
\midrule
AE1 & Token Efficiency & LLM Provider & add\_del\_ratio & +0.490 & 0.003 & +0.492 \\
AE2 & Token Efficiency & Extensibility & fix\_churn & +0.375 & 0.027 & +0.356 \\
AE3 & Token Efficiency & UI & add\_del\_ratio & +0.369 & 0.029 & +0.369 \\
AE4 & Token Efficiency & Context Mgmt & add\_del\_ratio & -0.346 & 0.042 & -0.392 \\
AE5 & Tool Calls Efficiency & LLM Provider & fix\_ratio & +0.367 & 0.030 & +0.373 \\
\bottomrule
\end{tabular}
\end{table}

\paragraph{Finding 14: LLM Provider and Extensibility associate with better token efficiency.}
Changes to the \textbf{Extensibility} layer correlate with improved token efficiency (AE2: $\rho=+0.375$, partial $\rho=+0.356$), as shown in Table~\ref{tab:rq3-ext-abs}. Expanding hook and plugin infrastructure allows the agent to solve tasks through specialized pathways rather than relying solely on general-purpose LLM reasoning, reducing the average token cost per resolution.

A similar pattern holds for the \textbf{LLM Provider} layer, as this layer shows the strongest association with token efficiency. This agent harness component is responsible for communicating with the LLM (i.e., formatting requests to and parsing responses from the LLM). When the LLM Provider changes are predominantly additions (\eg new model adapters, API expansions), the agent uses significantly fewer tokens (AE1: $\rho=+0.490$, partial $\rho=+0.492$). Bug fixes within the Provider also improve tool call efficiency (AE5: $\rho=+0.367$, partial $\rho=+0.373$). Additive changes expand capability without disrupting established token-efficient structures; in contrast, rewriting or deleting core interaction logic risks breaking tested pathways.

\paragraph{Finding 15: Context management expansion associates with lower token efficiency.}
Expanding \textbf{Context Management} correlates negatively with token efficiency (AE4: $\rho=-0.346$, partial $\rho=-0.392$), as shown in Table~\ref{tab:rq3-ext-abs}. Adding context window logic or scaling memory compaction increases the information presented to the LLM without improving reasoning quality. The strong partial correlation, which actually exceeds the raw value, confirms this is a genuine component effect rather than a release-size confound, suggesting that optimizing existing context structures is more beneficial than adding new ones.


\paragraph{Finding 16: UI refactoring shows divergent short- and long-term effects.}
The \textbf{UI} layer exhibits a notable split between the absolute and delta analyses. In the absolute analysis (Table~\ref{tab:rq3-ext-abs}), additive UI changes correlate with higher token consumption (AE3: $\rho=+0.369$, partial $\rho=+0.369$), suggesting that expanding user-facing features increases the agent's operational footprint over time. However, in the delta analysis (Table~\ref{tab:rq3-ext-delta}), UI refactoring correlates with improved effectiveness (DE1: $\rho=-0.391$, partial $\rho=-0.276$), indicating that restructuring existing UI code can yield immediate quality gains even if long-term UI expansion is costly.

\subsubsection{Delta ($\Delta$) Analysis: Immediate Release-to-Release Impacts}
While the absolute analysis captures long-term trends, it inherits accumulated state from prior releases. The delta analysis isolates the immediate impact of each release by measuring transitions between consecutive versions.

\begin{table}[t]
\centering
\caption{Extended Delta Analysis: immediate component-level transitions controlling for maturity, reporting only significant results ($p < 0.05$). Non-significant combinations are omitted. Row IDs (DE\#) uniquely identify each result for cross-referencing in the text (DE~=~Delta Effect). $\rho$~=~Spearman rank correlation; Partial~$\rho$~=~partial Spearman correlation controlling for total release churn; $p$~=~$p$-value.}

\label{tab:rq3-ext-delta}
\begin{tabular}{c l l l r r r}
\toprule
\# & Dimension & Component & Metric & Spearman $\rho$ & $p$ & Partial $\rho$ \\
\midrule
DE1 & Effectiveness & UI & refactor\_churn & -0.391 & 0.020 & -0.276 \\
DE2 & Effectiveness & LLM Provider & add\_del\_ratio & -0.336 & 0.049 & -0.221 \\
DE3 & Token efficiency & Persistence & add\_del\_ratio & +0.353 & 0.037 & +0.332 \\
DE4 & Token efficiency & Security & fix\_ratio & +0.346 & 0.042 & +0.327 \\
DE5 & Tool Calls Efficiency & Persistence & refactor\_churn & +0.345 & 0.042 & +0.334 \\
DE6 & Tool Calls Efficiency & Security & fix\_ratio & +0.341 & 0.045 & +0.339 \\
\bottomrule
\end{tabular}
\end{table}


\paragraph{Finding 17: Security modifications associate with efficiency gains.}
The \textbf{Security} domain stands out as a consistently safe target for modification. In the delta analysis (Table~\ref{tab:rq3-ext-delta}), fix-targeted security changes correlate with improved token efficiency (DE4: $\rho=+0.346$, partial $\rho=+0.327$) and reduced tool call overhead (DE6: $\rho=+0.341$, partial $\rho=+0.339$) without introducing the regressions observed in other components. A plausible explanation is that security fixes typically remove or tighten code paths rather than adding new ones, keeping the agent harness interaction surface compact and reducing the overhead introduced in each agent turn.

\paragraph{Finding 18: LLM Provider transitions associate with high regression risk; Persistence rework correlates with efficiency gains.}
Although the absolute analysis shows that additive LLM Provider changes are beneficial long-term, the delta analysis reveals that large LLM Provider modifications in a single release carry substantial immediate risk.
Releases with significant Provider changes show effectiveness degradation (DE2: $\rho=-0.336$, partial $\rho=-0.221$). Replacing model adapters or rewriting prompt formatting introduces untested code pathways that might degrade quality before the system stabilizes through subsequent patches.

The v0.4.1 to v0.5.0 transition exemplifies this risk. Despite relatively few PRs touching core agent logic, PR~\#1235 (``remove redundant if-check and add tests for OpenAI converter,'' +346/$-$8 lines) modified the Provider layer's request/response conversion pipeline, and PR~\#1245 removed an agent behavior mode (``corgi mode''). Resolve rate dropped from 39.4\% to 32.5\%, the lowest since v0.3.0, while token consumption remained elevated at 561K. The regression is subtle: the OpenAI converter change altered how the model's tool call requests are translated, a Provider-layer modification that is invisible to tool-level integration tests but directly affects the quality of the model's interactions.

Additive changes to the \textbf{Persistence Layer} (\eg updating session schemas or storage formats) are associated with improved token efficiency (DE3: $\rho=+0.353$, partial $\rho=+0.332$), while refactoring within Persistence is associated with tool call efficiency (DE5: $\rho=+0.345$, partial $\rho=+0.334$). A plausible explanation is that stabilized state management reduces the need for the LLM to re-process historical context redundantly across turns.


\begin{findingbox}[Summary of RQ3]
Architectural components differ substantially in their sensitivity to modification. The \textit{LLM Provider} layer carries significant immediate regression risk when modified extensively in a single release, as untested prompt pathways can degrade reasoning quality. \textit{Context Management} expansion increases prompt volume without improved effectiveness. Both require incremental, carefully staged modifications. On the other hand, \textit{Extensibility} changes correlate with improved token efficiency, while targeted \textit{Security} fixes consistently correlate with improved operational efficiency without introducing regressions. These represent the safest avenues for agent harness improvement.
\end{findingbox}


\section{Discussion}
\label{sec:discussion}

Here, we synthesize the findings from RQ0--RQ3, offer a likely explanation for the quality variations grounded in the absence of non-functional agentic regression testing, and present implications for practitioners, researchers, and the emerging field of agentic software engineering.

\subsection{More Code Does Not Mean Better Agents}
\label{sec:disc:paradox}

Across 35 \qwen~CLI releases, agent harness complexity increased steadily, token consumption nearly doubled, and tool call counts increased substantially, while agent effectiveness fluctuated across successive releases without a corresponding overall improvement (RQ1).
This pattern suggests that, given a particular agent harness design, ongoing development activity plays a central role in changing agent efficiency and effectiveness over time.
Later releases incurred higher execution costs without delivering better task outcomes. By holding the underlying LLM constant, we show that these trends arise from agent harness evolution rather than model changes.

Our results suggest that adding features, abstractions, and context-management logic does not necessarily improve agent quality. While these changes introduce new features and additional abstractions to expand the agent's capabilities, they also increase the amount of information processed by the model and the number of interactions with external tools. In our study, the additional overhead translated into higher token consumption and tool usage, but not into higher task resolution rates, and in some cases coincided with regressions.

This pattern is consistent with long-standing observations in software engineering that system growth tends to increase complexity faster than quality unless complexity is actively controlled~\cite{lehman1980programs}. In our case, token consumption and tool calls increased steadily across releases, whereas resolve rates remained largely stable.

These observations also have practical implications. Because API costs scale directly with token usage, the near-doubling of token consumption observed between early and late releases translates into a proportional increase in operational cost per task. For organizations deploying coding agents at scale, controlling agent harness complexity may therefore be as important as improving task effectiveness.

\subsection{The Absence of Non-functional Agentic Regression Testing}
\label{sec:disc:agentic-qa}

The regressions identified throughout RQ1--RQ3 were not covered by the project's automated testing infrastructure. Our inspection of the public CI/CD pipelines, test suites, and GitHub Actions workflows revealed functional testing, but no automated evaluation of agent-level benchmark quality metrics such as resolve rate, token consumption, or tool-call efficiency. We use the term \emph{non-functional agentic regression testing} to refer to automated evaluation of agent effectiveness and efficiency metrics, including resolve rate, token consumption, and tool calls across versions.

This gap provides a plausible explanation for why substantial quality regressions reached production releases. For example, the 139\% token increase in v0.1.5 and the 182\% increase in v0.3.0 both passed all automated checks. Similarly, the regressions associated with changes to the LLM Provider and Context Management layers (RQ3) were not detectable through existing tests because the CI pipeline evaluates whether the software functions correctly, not how effectively the agent performs its tasks. These regressions are quality regressions rather than functional defects. The affected releases passed functional tests and CI checks, yet still degraded agent effectiveness or efficiency.

The absence of agent-level evaluation was consistent throughout the \qwen~CLI repository. While the project maintains more than 300 unit tests, integration tests, and end-to-end tests, none evaluate the agent on representative coding benchmarks or enforce budgets on token consumption and tool usage. This is likely attributable to the cost of running agentic evaluations at the commit cadence of an actively developed project, rather than a lack of awareness of the importance of such testing.

Taken together, these observations suggest that the development process lacks feedback mechanisms for detecting degradations in agent effectiveness and efficiency. We do not claim that the absence of non-functional agentic regression testing is the sole cause of the quality drops. However, the regressions documented in this study could not have been detected by the project's public testing infrastructure before release. This highlights the need for evaluation pipelines that track agent-level metrics across versions and flag regressions in effectiveness and efficiency.

\subsection{Agentic CLIs as a New Category of Software}
\label{sec:disc:new-category}

Our findings suggest that agentic CLIs represent an emerging category of software that challenges existing quality assurance practices. Unlike traditional tools, where functional correctness is often the primary quality criterion, coding agents must be evaluated across multiple dimensions.

The results of this study show that these dimensions can evolve independently. Across multiple releases, the \qwen~CLI passed automated tests, yet resolve rates, token consumption, and tool call counts changed substantially. This is a consequence of the probabilistic, open-ended nature of agentic behavior, which binary pass/fail tests are not designed to capture.
A release may therefore be functionally correct while becoming more expensive to operate or less effective at solving tasks. Quality in agentic systems extends beyond correctness and includes task effectiveness and resource efficiency.

This distinction stems from the fact that coding agents operate through interactions between agent harness components and an LLM. Changes to prompts, context management, or tool interactions can alter agent behavior even when the underlying model remains unchanged. As a result, evaluating coding agents requires quality assurance practices that complement traditional testing with agent-level evaluation. Studying and developing such practices represents an important direction for future software engineering research.

These observations motivate a form of quality assurance tailored to coding agents. In addition to functional testing, Agentic QA should consider non-functional agent-level outcomes such as task effectiveness and resource efficiency across releases. Such evaluation can help identify regressions that remain invisible to traditional functional tests.

\section{Threats to Validity}
\label{sec:threats}

\textbf{Internal Validity.}
LLM stochasticity is a known confounding factor in agent evaluations. To mitigate this, we used the default sampling parameters of the \qwen~CLI and executed each task twice per release. The resulting run-to-run agreement rate of 87.7\% (Section~\ref{sec:rq1:results}), where both runs were strongly consistent in task outcomes, confirms that observed quality variations are attributable to agent harness changes rather than model non-determinism. By self-hosting the LLM via vLLM on dedicated hardware, we also eliminated external API variability (timeouts, rate-limiting, silent model updates) as a confound.

Our statistical analyses in RQ2 and RQ3 employ multiple testing correction (Benjamini-Hochberg) to control the false discovery rate. However, the sample size of 35 releases limits statistical power, particularly for the delta analysis where $n=34$. We report effect sizes (Cliff's $d$) alongside significance tests to ensure that reported findings reflect practically meaningful differences.

\textbf{External Validity.}
The primary threat to external validity is our focus on a single coding agent harness (the \qwen~CLI) for the controlled longitudinal analysis (i.e., RQ1--3). Whether the results generalize to other agent harnesses such as SWE-agent, OpenHands, or Codex remains an open question. We deliberately prioritized longitudinal depth (35 releases $\times$ 50 tasks $\times$ 2 runs = 3,500 inference executions and evaluations), which required several weeks of computation, over multi-harness breadth, as this design enables the controlled experimentation required to isolate the agent harness variable. The broad analysis in RQ0 provides complementary evidence that the hyper-churn development patterns we observe are industry-wide, suggesting that the conditions for quality fluctuations are present across multiple agent harnesses. Nevertheless, we encourage future replication studies across additional agent implementations.

Our RQ0 analysis relies on GitHub repository data (commits, PRs, issues), which carries known threats when mining software repositories~\cite{kalliamvakou2014promises}. Not all development activity is visible through the public repository; some changes may occur in private forks or internal branches before being merged.

Our evaluation uses a single LLM (Qwen3-Next-80B-A3B-Instruct). We selected this model because \qwen~is designed and optimized for Qwen-series models,\footnote{\url{https://github.com/QwenLM/qwen-code}} and because \qwen~is a strong open-weight model that can be self-hosted via vLLM on dedicated hardware, eliminating external API variability as a confound. Different models may interact with agent harness changes differently. A model with stronger instruction-following capabilities might be more robust to prompt bloat, while a weaker model might be more sensitive. Our findings should be interpreted as evidence that agent harness matters, not as universal quantitative predictions for all model-harness combinations.

The task sample (50 tasks from \swebench~Verified) represents a subset of the full benchmark. While our stratified sampling preserves the difficulty distribution, the smaller sample size may not capture rare task types where agent harness changes have outsized effects.

Our selection criteria for RQ0 (open-source, actively maintained, CLI-based) ensure that the projects under study have sufficient release history to observe meaningful evolution.
Agent harnesses with few or no releases would not meet these criteria, which means that our characterization of high development activity may not generalize to less active projects. However, our goal in RQ0 is precisely to characterize the most actively developed agent harnesses, since these are the ones most likely to exhibit the evolution patterns we study.

\textbf{Construct Validity.}
We use the \swebench~Resolve Rate as our primary effectiveness metric. This binary metric (all tests pass or not) does not capture partial progress: a patch that fixes the core logic but fails a single edge-case test is scored identically to a patch that makes no meaningful changes. However, Resolve Rate is the standard metric in the \swebench~ecosystem and enables direct comparison with prior work. Future studies could supplement it with partial-credit metrics (\eg percentage of tests passed) to capture finer-grained quality differences.

Our stratified sample of 50 tasks was drawn to represent the difficulty distribution of \swebench~Verified. However, task difficulty is estimated using fix time (issue creation to patch merge), which is an imperfect proxy; some fast fixes involve deep reasoning, while some slow fixes reflect process delays rather than inherent difficulty.

The classification of commits into architectural components (RQ3) involves manual judgment in mapping files to the reference architecture. 
While we mapped files to components through systematic manual inspection of the codebase guided by our reference architecture (Appendix~\ref{sec:architecture}), and documented each assignment decision explicitly, some files span multiple components and required proportional allocation. Different mapping decisions could yield slightly different results at the component level, though the high-level patterns are robust to reasonable alternative mappings.

\section{Related Work}
\label{sec:related}

Our work sits at the intersection of three research areas: coding agents and their evaluation, software evolution and release engineering, and empirical studies of AI-assisted development. We discuss each in turn and position our contributions relative to the existing literature.

\subsection{Coding Agents and Benchmarks}
\label{sec:related:agents}

The development of autonomous coding agents has accelerated rapidly. SWE-agent~\cite{yang2024sweagent} introduced an agent-computer interface designed to help LLMs navigate and edit codebases, achieving strong results on \swebench. OpenHands~\cite{wang2024openhands} provides an open platform for building generalist AI software developers. These systems, along with others such as AutoCodeRover~\cite{zhang2024autocoderover} and Agentless~\cite{xia2024agentless}, represent different points in the design space of coding agents, from heavily scaffolded agentic loops to lightweight, non-agentic approaches.

The \swebench~benchmark~\cite{jimenez2024swebench} has become the standard for evaluating coding agents on real-world software engineering tasks. It provides a collection of GitHub issues paired with test-validated ground-truth patches, enabling reproducible evaluation of agent capabilities. The curated \swebench~Verified subset further improves evaluation reliability by filtering for well-specified tasks.

A key limitation of existing benchmarking work is that evaluations are typically performed as \emph{static snapshots}: a single version of an agent is evaluated once, and the result is reported as representative. This approach masks the substantial version-to-version quality variation that we document in this study. 
Furthermore, existing evaluations rarely control for the agent harness version (i.e., reporting results for a single agent harness version without tracking how quality evolves across versions). This makes it impossible to attribute quality changes to the agent harness versus the LLM. Our study addresses both limitations by performing a controlled longitudinal evaluation across 35 releases while holding the LLM constant.

Li~\etal~\cite{li2025aidev} recently introduced the AIDev dataset, capturing over 456,000 pull requests authored by five leading coding agents. Their work provides empirical evidence that coding agents are active participants in real-world software development, but focuses on the \emph{outputs} of agents (pull requests) rather than the \emph{evolution} of the agent systems themselves. Hassan~\etal~\cite{hassan2025sase} articulated the vision of Structured Agentic Software Engineering (SASE), arguing for a systematic rethinking of SE practices in the agentic era. Our work complements both studies by empirically examining the internal evolution of the agent harness layer that powers these agents.

\subsection{Software Evolution and Release Engineering}
\label{sec:related:evolution}

The study of software evolution has a long tradition in software engineering. Lehman's laws of software evolution~\cite{lehman1980programs} established that actively used software systems must continuously adapt to remain satisfactory, yet this continuous change tends to increase complexity and degrade quality unless explicitly counteracted. Parnas~\cite{parnas1994software} formalized the concept of ``software aging,'' arguing that software deteriorates over time due to both the failure to adapt and the accumulation of poorly planned changes. Empirical studies of open-source evolution, from the Linux kernel~\cite{godfrey2000evolution} to Apache and Mozilla~\cite{mockus2002two}, have confirmed these patterns in large-scale systems. These classical observations are directly relevant to our findings: coding agent harness evolves at an extreme pace, and our results suggest that this rapid evolution does not consistently translate into improved agent quality.

Code churn, i.e., the volume of code added, deleted, or modified over time, has been shown to be a strong predictor of software defects~\cite{nagappan2005churn}. Hindle~\etal~\cite{hindle2008large} developed a taxonomy of large commits, showing that commit size and composition carry distinct implications for software quality. The pull-based development model~\cite{gousios2014pullbased}, now dominant in open-source projects, introduces additional dynamics through code review workflows and contributor coordination. Our study extends these lines of work to the domain of coding agents, examining whether the volume and nature of agent harness changes correlate with quality regressions.

Release engineering research has examined how release practices affect software quality. Studies of rapid release cycles in projects like Firefox and Chrome have shown that faster release cadences can improve time-to-market but may introduce quality trade-offs~\cite{adams2016release, khomh2015rapid, khomh2012faster}. The release velocities we observe in coding agent harness (exceeding two releases per day in some projects) are substantially higher than those reported in prior release engineering studies, suggesting that coding agent harnesses represent a qualitatively different development regime.

Technical debt~\cite{kruchten2012technical} provides another lens through which to understand harness evolution. The continuous accumulation of hastily implemented features and deferred maintenance can lead to a system that becomes increasingly difficult to modify without introducing regressions. Our architectural sensitivity analysis (RQ3) provides empirical evidence for this phenomenon in the specific context of coding agents.

\subsection{Empirical Studies of AI-Assisted Development}
\label{sec:related:empirical}

A growing body of work examines the effectiveness of AI tools in software development. Recent surveys~\cite{hou2024llm4se, liu2024llmagent} systematically catalogue the landscape of LLM applications in software engineering. Studies of GitHub Copilot~\cite{peng2023copilot} have measured productivity gains from AI-assisted coding. Evaluations of code LLMs on benchmarks like HumanEval~\cite{chen2021codex} have tracked improvements in raw code generation capability. However, these studies focus on the \emph{model's} capability rather than the \emph{agent harness} contribution to agent quality.

The distinction between model capability and agent harness effectiveness is critical, yet routinely overlooked. When a coding agent improves or regresses between two releases, practitioners commonly attribute the change to the underlying LLM, fueling speculation about model updates or silent capability changes, while ignoring modifications to the agent harness layer (system prompts, tool routing logic, context management strategies) that are invisible to standard benchmarking methodologies. Our study is the first to empirically isolate and quantify the agent harness contribution to quality by holding the model constant across all 35 releases.

The ReAct framework~\cite{yao2023react} demonstrated that interleaving reasoning traces with action steps significantly improves LLM performance on interactive tasks, building on the chain-of-thought prompting paradigm~\cite{wei2022chain}. This pattern (reason, then act, then observe) has become the foundational loop in most coding agent harnesses. Our reference architecture captures this pattern in the Orchestrator component, and our analysis reveals that modifications to how this loop is implemented can have substantial effects on agent quality.

Our work complements these efforts by focusing on a dimension that has not been empirically explored: the longitudinal relationship between agent harness evolution and agent quality, studied under controlled conditions that isolate the agent harness variable. While prior work benchmarks agent capabilities at a single point in time or characterizes agent outputs, we track how quality changes across 35 sequential releases and explain these shifts at two complementary levels: at the project level, through development patterns such as code churn, commit types, and contributor activity (RQ2); and at the architecture level, by mapping changes to specific structural components of a reference architecture derived from five agent harnesses (RQ3).

\section{Conclusion}
\label{sec:conclusion}

Coding agent harnesses are evolving at an unprecedented pace, with new features and architectural changes introduced to improve the agent quality (i.e., effectiveness and efficiency). This paper provides a controlled empirical evaluation of how this evolution affects effectiveness and efficiency over time, while holding the LLM version constant.

We first characterized the rapid development activity across five major coding agent harnesses (RQ0), revealing release frequencies, contributor turnover, and issue volumes that substantially exceed those of mature open-source projects. We then performed the first controlled, longitudinal evaluation of agent harness evolution (RQ1), evaluating 35 sequential releases of the \qwen~CLI on 50 \swebench~Verified tasks while holding the underlying LLM version constant. This revealed that later releases consume nearly double the tokens and tool calls of earlier versions while achieving no statistically significant improvement in resolve rates.

We explained these quality shifts at two complementary levels. At the \emph{project level} (RQ2), feature additions were the only catalyst for improved resolve rates, but they came at the cost of increased resource consumption. In contrast, fix-heavy releases, scattered small changes, and new contributor onboarding increased costs without improving effectiveness. At the \emph{architecture level} (RQ3), we identified the LLM Provider layer and Context Management as high-risk zones where modifications carry the greatest regression risk, while Extensibility investments and targeted Security fixes consistently yielded safe improvements. Qualitative analyses further linked specific architectural changes to measurable quality regressions.

These findings highlight the need for \emph{Non-functional Agentic Quality Assurance}, specifically regression testing that evaluates agent effectiveness and efficiency across releases, a practice well established for traditional software systems but not yet adopted for agentic systems, likely due to the high cost of running representative evaluations at the commit cadence of an actively developed project. As agentic systems grow in complexity, developers should monitor metrics such as resolve rate, token consumption, and tool call overhead across releases, while researchers should report and control for agent harness versions alongside the underlying LLM.

\bibliography{bib/references}

\clearpage
\appendix
\section{Appendix: Reference Architecture for Coding Agent Harness}
\label{sec:architecture}

Before mapping code changes to architectural components in RQ3, we require a common vocabulary and a structural decomposition of coding agent harness. 

A \emph{reference architecture} is an architecture template for all software systems in a domain~\cite{hassan2000reference}. It defines the fundamental components of the domain and the relations between them. The architecture for a particular product is an instance of the reference architecture. During development, the product designer refines and extends the reference architecture based on the product's requirements and constraints. Reference architectures provide a common nomenclature across systems, enable uniform comparison of different products, and facilitate both forward and reverse engineering~\cite{hassan2000reference}.

Since no such reference architecture exists for this emerging domain, we derive one following the process established by Hassan and Holt~\cite{hassan2000reference}. We first derive a conceptual architecture for each agent harness from its documentation, map it to a concrete architecture by tracing actual code components, identify commonalities across the five concrete architectures, and synthesize a unified reference architecture that captures the shared structure.

\subsection{Derivation Process}
\label{sec:architecture:process}

Our derivation follows the two-step process of Hassan and Holt~\cite{hassan2000reference}, adapted for the coding agent domain:




\paragraph{Step~1: Conceptual and Concrete Architecture Extraction.} 
For each of the five agent harnesses in our study (i.e., Codex, OpenCode, Gemini~CLI, Qwen~Code, and OpenHands~CLI), we first derive a \emph{conceptual architecture} from each project's documentation (e.g., READMEs, design docs, architecture diagrams) and publicly available discussions, identifying major subsystems, their responsibilities, and inter-subsystem relationships at an abstract level, without consulting the source code. We then map this conceptual architecture onto a \emph{concrete architecture} by manually analyzing the source code, tracing each conceptual subsystem to its implementing source files, classes, and entry points\footnote{Source code analyzed as of March 2026.}. We iteratively refine both representations until they align. For each agent harness, we:

\begin{enumerate}[leftmargin=*]
    \item Examine directory structures, module boundaries, and package organization to identify major subsystems and their conceptual roles.
    \item Trace module graphs and function call chains to recover inter-subsystem dependencies and data flow, mapping each conceptual subsystem to its concrete code elements.
    \item Identify the communication mechanisms (typed channels, event buses, observer patterns) that connect subsystems.
    \item Document each subsystem's responsibilities, primary classes, and entry points.
\end{enumerate}

This analysis covered five codebases spanning four programming languages (Rust, TypeScript, Python, Java) and hundreds of source files.
The result is a concrete architecture for each agent harness, documented as written descriptions of each subsystem's responsibilities and relationships (Appendix~\ref{sec:architecture}), and operationalized as a file-to-component mapping that assigns every source file to exactly one architectural component, available in the replication package.

\paragraph{Step~2: Reference Architecture Synthesis.} Using the five concrete architectures from Step~1, we identify common components that recur across implementations and abstract them into a reference architecture following an iterative refinement process. Concretely, we:

\begin{enumerate}[leftmargin=*]
    \item[\textbf{2a.}] \emph{Propose} a candidate reference architecture based on domain knowledge and the most frequently occurring subsystems across the five concrete architectures.
    \item[\textbf{2b.}] \emph{Refine} the candidate by validating it against each concrete architecture: verify that every reference component maps to at least one subsystem in each agent harness, and revise the boundaries to accommodate implementation variation (splitting, merging, or renaming of subsystems).
\end{enumerate}

Components that appear in only one harness (e.g., OpenCode's database interface subsystem) are excluded from the reference architecture but noted as implementation-specific extensions. The final reference architecture represents the structural consensus across all five agent harnesses.


\subsection{The Reference Architecture of AI coding agent harness}
\label{sec:architecture:reference}

\begin{figure}[t]
\centering
\includegraphics[width=\textwidth]{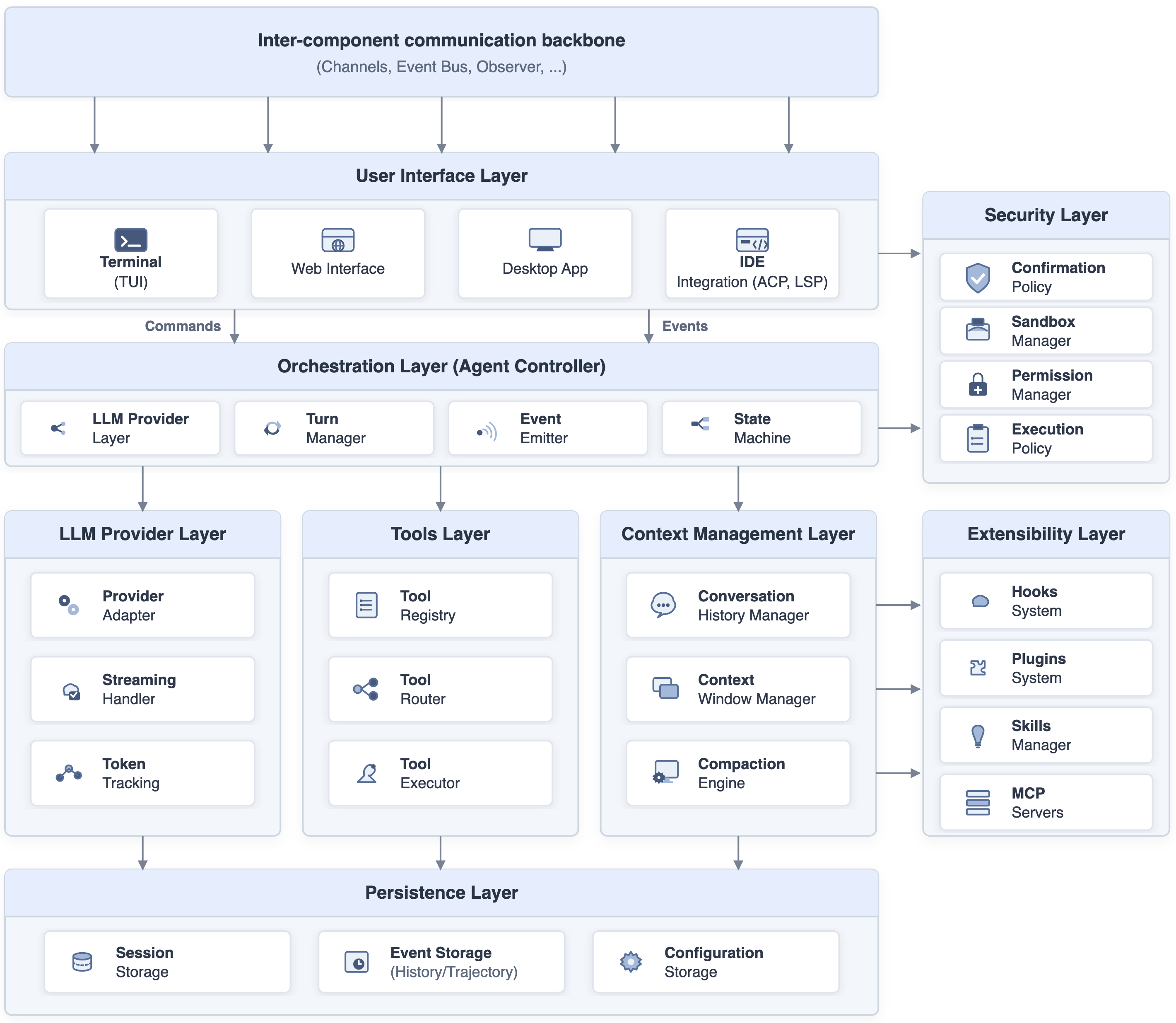}
\caption{Reference architecture for AI coding agent harness. The Communication Backbone (top) connects all components. Core components (center) implement the agent's primary functionality. Cross-cutting components (sides) provide infrastructure services. Arrows indicate communication flow.}
\label{fig:reference-architecture}
\end{figure}

Figure~\ref{fig:reference-architecture} presents the reference architecture. It comprises ten components organized into three tiers: a \emph{communication backbone} that connects all components, \emph{core components} that implement the agent's primary functionality, and \emph{cross-cutting components} that provide infrastructure services to multiple core components.

The architecture is \emph{not} a simple layered or sequential structure. Components have complex interconnections with bidirectional communication, event-driven messaging, and cross-cutting concerns. We now describe each component.

\subsubsection{Communication Backbone}
\label{sec:architecture:backbone}

The \emph{Inter-Component Communication Backbone} is the conceptual nervous system of the architecture. It enables all components to communicate without tight coupling. The backbone is not a layer above other components; rather, it is a cross-cutting infrastructure that all components connect to. Three implementation strategies emerge across the five agent harnesses:

\begin{itemize}[leftmargin=*]
    \item \textbf{Typed Async Channels} (Codex): Point-to-point communication with compile-time type checking, where each channel connects exactly one sender to one receiver.
    \item \textbf{Event Bus / Pub-Sub} (Gemini~CLI, Qwen~Code, OpenCode): Topic-based publish-subscribe with decoupled publishers and multiple subscribers. Subscriptions can be added or removed at runtime.
    \item \textbf{Observer Pattern} (OpenHands~CLI): The SDK pushes events to registered observer callbacks in a single direction. The CLI wraps the SDK and bridges blocking calls to async patterns.
\end{itemize}

The choice of backbone mechanism is strongly influenced by the implementation language: Rust's type system enables compile-time-safe channels, TypeScript agent harnesses favor event emitters, and Python uses callback-based observers.

\subsubsection{Core Components}

\paragraph{1. User Interface (UI) Layer.} Handles all user interaction and presentation. Communication with the backend is \emph{bidirectional}: events flow up from the agent (streaming text, tool results, status changes, confirmation requests), while commands flow down from the user (prompts, approvals, interrupts). All five agent harnesses implement a terminal-based UI (TUI) as the primary interface, though each uses a different framework: ratatui in Rust (Codex), React/Ink in TypeScript (Gemini~CLI, Qwen~Code), SolidJS (OpenCode), and Textual in Python (OpenHands~CLI). Several agent harnesses additionally provide IDE integrations via protocols such as ACP (Agent Communication Protocol), LSP, and MCP.

\paragraph{2. Agent Controller (Orchestrator).} The central coordinator that manages the agent's lifecycle and implements the core agent loop: \emph{prompt} $\rightarrow$ \emph{LLM} $\rightarrow$ \emph{tool} $\rightarrow$ \emph{response}. This component realizes the ReAct reasoning cycle~\cite{yao2023react}. It contains four sub-elements: a \emph{Session Manager} that manages conversation state and session lifecycle; a \emph{Turn Manager} that handles individual conversation turns (building context, sending requests to the LLM, coordinating tool execution); an \emph{Event Emitter} that broadcasts events to connected components; and a \emph{State Machine} that tracks agent execution states (Idle $\rightarrow$ Executing $\rightarrow$ Waiting for Confirmation / Paused / Finished / Error).

\paragraph{3. LLM Provider Layer.} Abstracts communication with the underlying language model. It consists of three sub-elements: a \emph{Provider Adapter} that normalizes different LLM APIs into a common interface; a \emph{Streaming Handler} that processes Server-Sent Events (SSE), buffers content deltas, and handles tool call streaming; and a \emph{Token Counter} that tracks input/output token usage and enforces model-specific context limits. The number of supported providers varies significantly: OpenCode supports 22 providers, Qwen~Code supports 7 (DashScope, OpenAI, Anthropic, Gemini, DeepSeek, OpenRouter, ModelScope), Codex supports 3, and Gemini~CLI supports only Google Gemini natively.

\paragraph{4. Tool System Layer.} Manages the registration, routing, and execution of tools, i.e., the mechanisms through which the agent takes actions in the environment (reading files, executing commands, searching the web). It consists of a \emph{Tool Registry} that manages tool definitions and provides schemas to the LLM; a \emph{Tool Router} that dispatches tool calls from LLM responses to the appropriate handler; and a \emph{Tool Executor} that runs tools, potentially in sandboxed environments, and formats results for LLM consumption. Tool counts vary by an order of magnitude across coding agent harnesses (Gemini~CLI: 34, OpenCode: 25, Codex: 23, Qwen~Code: 17, OpenHands~CLI: 4 default), though all five support MCP for tool extensibility.

\subsubsection{Context and Data Components}

\paragraph{5. Context Management Layer.} Responsible for assembling, compressing, and curating the information presented to the LLM at each turn. It contains a \emph{Conversation History Manager} that tracks the conversation state (some agent harnesses distinguish between ``curated'' and ``comprehensive'' history); a \emph{Context Window Manager} that monitors token usage against model limits; and a \emph{Compaction Engine} that compresses context when limits are approached. All five agent harnesses implement context compaction, reflecting the universal challenge of managing finite context windows.

\paragraph{6. Persistence Layer.} Handles data storage for sessions, event histories, and configuration. Persistence models differ across coding agent harnesses: Codex and Qwen~Code use append-only JSONL files; OpenCode uses SQLite with an ORM (Drizzle); Gemini~CLI uses JSON files via a dedicated recording service; and OpenHands~CLI uses an event-sourcing model. The Persistence Layer is accessed by the Session Manager, Context Manager, and Tool System for writing histories and audit logs, and by the Config component for reading stored settings.

\subsubsection{Cross-Cutting Components}

The remaining four components provide infrastructure services that interact with multiple core components. They are \emph{not} sequential layers; each connects independently to the components it serves.

\paragraph{7. Security Layer.} Enforces security policies for tool execution, including tool confirmation, sandboxing, and permission management. This component shows the widest implementation variation across agent harnesses. It consists of a \emph{Confirmation Policy} that determines which tool calls require user approval; a \emph{Sandbox Manager} that provides isolated execution environments; a \emph{Permission Manager} that tracks allowed and denied operations; and an \emph{Execution Policy} that enforces constraints on what the agent can do. Codex implements the most comprehensive sandboxing with cross-platform support (Seatbelt on macOS, bubblewrap on Linux, Landlock, Windows containers). Gemini~CLI uses a separate PolicyEngine service with Seatbelt and Docker. Qwen~Code replaces the policy engine with a hook-based PermissionRequest system. OpenCode embeds permissions in the agent logic without runtime sandboxing. OpenHands~CLI implements a strategy pattern for confirmation policies without sandboxing in the CLI.

\paragraph{8. Extensibility Layer.} Enables the agent to be extended through hooks, plugins, skills, and external tool servers. It consists of a \emph{Hooks System} that provides lifecycle event interception (e.g., pre/post tool use, session start/end); a \emph{Plugins/Skills System} that enables user-defined capabilities; and \emph{MCP Server} support for external tool integration. Qwen~Code and Gemini~CLI have the most comprehensive hook systems (12 and 11 event types respectively). Qwen~Code uniquely implements an extension marketplace for community-developed extensions.

\paragraph{9. Config / Service Locator.} Provides configuration management and dependency injection. This component is accessed by all other components for configuration values and service instances. Implementations range from Rust module-based configuration (Codex) to TypeScript service locator patterns with hierarchical configuration resolution (Gemini~CLI, Qwen~Code), a DI container with lazy initialization (OpenCode), and a Python agent store (OpenHands~CLI).


\subsection{Mapping to Concrete Architectures}
\label{sec:architecture:mapping}

Following Hassan and Holt~\cite{hassan2000reference}, we validate the reference architecture by mapping it to the concrete architecture of each agent harness. For each agent harness, we provide a brief characterization and highlight notable deviations from the reference architecture.

\subsubsection{Codex}

Codex's architecture is organized around a hub-and-spoke pattern: the central \texttt{Codex} struct uses typed channels (\texttt{Sender<Submission>}, \texttt{Receiver<Event>}) for point-to-point communication with bounded backpressure (i.e., a fixed-size buffer between sender and receiver prevents message accumulation by blocking the sender when the buffer is full). The orchestrator coordinates a \texttt{ContextManager}, \texttt{ModelClient}, and \texttt{ToolRouter}. Codex implements the most comprehensive security model, with cross-platform sandboxing via Seatbelt (macOS), bubblewrap (Linux), Landlock, and Windows containers. Notable deviations: Codex merges the Context Management and Session Manager into a single \texttt{SessionState} struct, whereas the reference architecture treats them as separate components. Its Extensibility Layer is minimal: it lacks a comprehensive hook system (no lifecycle event interception), unlike Gemini~CLI and Qwen~Code.

\subsubsection{OpenCode}

OpenCode's architecture follows a dependency injection pattern: a central \texttt{Instance} DI container provides lazy-initialized services. The orchestrator (\texttt{SessionPrompt.loop()}) communicates via an \texttt{EventEmitter}-based event bus. OpenCode uniquely uses SQLite with the Drizzle ORM for persistence, whereas all other agent harnesses use file-based storage. Notable deviation: the Security Layer lacks runtime sandboxing entirely; security is embedded directly in the agent logic as permission checks rather than implemented through a separate sandbox or policy engine subsystem. This represents a reduced instantiation of the Security component rather than its absence.

\subsubsection{Gemini CLI}

Gemini~CLI's architecture follows a service locator pattern: a central \texttt{Config} object provides access to services including the \texttt{MessageBus}, \texttt{GeminiClient}, and \texttt{PolicyEngine}. The orchestrator (\texttt{GeminiClient}) uses a streaming content generator pattern. Gemini~CLI maintains two parallel histories (a ``curated'' history of valid conversation turns and a ``comprehensive'' full message history) reflecting a unique approach to context management. Notable deviation: the LLM Provider Layer supports only a single provider (Google Gemini). The Provider Adapter sub-element is effectively absent; the agent harness communicates directly with the Gemini API without an abstraction layer for alternative providers.

\subsubsection{Qwen Code}

Qwen~Code shares a common heritage with Gemini~CLI, from which it was forked. Its architecture retains the service locator pattern and \texttt{MessageBus} communication, but diverges significantly in three areas: (1)~the LLM Provider Layer supports seven providers (Qwen/DashScope, OpenAI, Anthropic, Gemini, DeepSeek, OpenRouter, ModelScope), each implemented as a separate provider adapter, compared to Gemini~CLI's single-provider design; (2)~the Security Layer replaces Gemini~CLI's \texttt{PolicyEngine} with a hook-based \texttt{PermissionRequest} system that auto-approves tool confirmations via the MessageBus; and (3)~the Extensibility Layer adds a unique extension marketplace. Qwen~Code's hook system (12 event types) is the most comprehensive among the five agent harnesses. Notable deviation: some files span multiple reference components (e.g., \texttt{sharedTokenManager.ts} belongs to both the LLM Provider and Context Management components), reflecting cross-component responsibilities.

\subsubsection{OpenHands CLI}

Unlike the other four agent harnesses, OpenHands~CLI is a thin CLI wrapper around the OpenHands SDK rather than a standalone agent. Its architecture follows an adapter pattern: the CLI wraps the SDK's \texttt{BaseConversation} and bridges blocking SDK calls to async patterns using \texttt{asyncio.to\_thread}. Communication uses an observer pattern (\texttt{EventSubscriber}) where the SDK pushes events to registered callbacks. The CLI delegates most core functionality (LLM interaction, tool execution, context management) to the SDK, implementing only the UI, confirmation workflow, and persistence locally. Notable deviations: the Security Layer has no runtime sandboxing in the CLI (though the SDK supports Docker-based sandboxing); the Extensibility Layer is minimal, lacking both a comprehensive hook system and a skills system; and the Tool System has only 4 default tools, relying heavily on MCP for extensibility.

\subsubsection{Summary of Mappings}

Table~\ref{tab:arch-mapping} summarizes the mapping of reference components to concrete implementations across all five agent harnesses. The mapping confirms that all ten top-level components have corresponding implementations in each agent harness, validating the reference architecture's completeness. However, not all \emph{sub-elements} within components are universally present. These gaps represent reduced instantiations rather than entirely missing components. Most notably, the Security Layer varies the most: OpenCode and OpenHands~CLI lack runtime sandboxing, while Codex provides cross-platform process isolation. Similarly, the Extensibility Layer ranges from minimal (Codex, OpenHands~CLI) to comprehensive (Qwen~Code with 12 hook types and an extension marketplace). Gemini~CLI's LLM Provider Layer lacks a provider abstraction, communicating directly with the Gemini API. Despite these variations, the agent harnesses, developed by five different organizations, in four programming languages, with distinct design goals, all fit the common reference architecture. The main structural differences are the splitting and merging of subsystems (e.g., Codex merges Context Management into the Session Manager; Gemini~CLI splits history into curated and comprehensive) and differing numbers of sub-elements within components.

\begin{table*}[t]
\caption{Mapping of reference architecture components to concrete implementations. Each cell shows the primary implementation artifact.}
\label{tab:arch-mapping}
\centering
\resizebox{\textwidth}{!}{%
\begin{tabular}{l l l l l l}
\toprule
\textbf{Reference Component} & \textbf{Codex} & \textbf{OpenCode} & \textbf{Gemini CLI} & \textbf{Qwen Code} & \textbf{OpenHands} \\
\midrule
Comm.\ Backbone         & Typed Channels             & Event Bus (\texttt{EventEmitter}) & MessageBus (pub/sub)        & MessageBus (auto-approve)       & Observer (\texttt{EventSubscriber}) \\
User Interface           & ratatui TUI                & SolidJS Console                  & React/Ink TUI               & React/Ink TUI + i18n            & Textual TUI \\
Orchestrator             & \texttt{Codex} struct       & \texttt{SessionPrompt.loop()}    & \texttt{GeminiClient}       & Content Generators + Client     & \texttt{ConversationRunner} \\
LLM Provider             & Multi (3 providers)         & 22 providers                     & Gemini only                 & 7 providers (DashScope, etc.)   & Multi (via SDK + LiteLLM) \\
Tool System              & 23 tools                    & 25 tools                         & 34 tools                    & 17 tools                        & 4 default + MCP \\
Context Mgmt             & \texttt{ContextManager}     & Compaction service               & Curated + comprehensive     & Content gen.\ integrated        & SDK-managed \\
Security                 & Sandbox (cross-platform)    & Embedded permissions             & PolicyEngine + Seatbelt     & PermissionRequest hooks         & Strategy pattern \\
Persistence              & JSONL (append-only)         & SQLite (Drizzle ORM)             & JSON files (recording svc.) & JSONL (append-only)             & Event-sourcing \\
Extensibility            & Limited hooks               & 7 hook types                     & 11 event types              & 12 events + marketplace         & Limited \\
Config / Svc.\ Locator   & Module config               & DI Container (\texttt{Instance}) & Service Locator (\texttt{Config}) & Svc.\ Locator + ModelRegistry  & Agent Store \\
\bottomrule
\end{tabular}%
}
\end{table*}

The number of conceptual subsystems recovered from each agent harness ranges from 8 (OpenHands~CLI) to 13 (Qwen~Code), reflecting differences in codebase scope and modularity. Table~\ref{tab:rq0-subjects} in Section~\ref{sec:rq0} provides additional statistics for each agent harness.


\end{document}